\documentclass[%
 reprint,
twocolumn,
superscriptaddress,
 amsmath,amssymb,
 aps,
 prc, 
]{revtex4-2}

\usepackage{graphicx}
\usepackage{dcolumn}
\usepackage{bm}
\usepackage{color}
\newcommand{\epg}{e p \to e p \gamma}
\newcommand{\epgreactnucleon}{e N \to e N \gamma}

\newcommand{\rcsreactnucleon}{\gamma N \to N \gamma}

\newcommand{\vcsreactnucleon}{\gamma^* N \to N \gamma}

\newcommand{\qcm}{q_\mathrm{c.m.}}
\newcommand{\qpr}{q'_\mathrm{c.m.}}
\newcommand{\thcm}{\theta_{\mathrm{c.m.}}}
\newcommand{\phicm}{\phi_{\mathrm{c.m.}}}
\newcommand{\cthcm}{\cos \theta_{\mathrm{c.m.}}}
\newcommand{\mxx}{M_X^2}
\newcommand{\zve}{Z_{\mathrm{vertex}}}
\newcommand{\xve}{X_{\mathrm{vertex}}}
\newcommand{\yve}{Y_{\mathrm{vertex}}}
\newcommand{\ybeam}{Y_{\mathrm{beam}}}
\newcommand{\xbeam}{X_{\mathrm{beam}}}
\newcommand{\ebeam}{E_{\mathrm{beam}}}
\newcommand{\ztarget}{Z_{\mathrm{target}}}
\newcommand{\efrost}{e_{\mathrm{frost}}}
\newcommand{\pllptte}{P_{LL} -P_{TT} / \epsilon}
\newcommand{\plt}{P_{LT}}
\newcommand{\ptt}{P_{TT}}
\newcommand{\pll}{P_{LL}} 
\newcommand{\la}{\Lambda_{\alpha}}
\newcommand{\lb}{\Lambda_{\beta}}
\newcommand{\aeq}{\alpha_{E1}(Q^2)}
\newcommand{\bmq}{\beta_{M1}(Q^2)}

\newcommand{\ohigher}{{\cal O}(q_\mathrm{c.m.}^{\prime 2})}
\newcommand{\ohigherdr}{{\cal O}(q_\mathrm{c.m.}^{\prime 2})_{\mathrm{DR}}}
\newcommand{\sigbhb}{d \sigma_{\mathrm{BH+Born}}}

\newcommand{\sigexp}{d \sigma_{\mathrm{exp}}} 
\newcommand{\sigdr}{d \sigma_{\mathrm{DR}}}
\newcommand{\siglex}{d \sigma_{\mathrm{LEX}}}

\newcommand{\xfp}{x_{\mathrm{fp}}}
\newcommand{\yfp}{y_{\mathrm{fp}}}
\newcommand{\thfp}{\theta_{\mathrm{fp}}}
\newcommand{\phfp}{\phi_{\mathrm{fp}}}
\newcommand{\xfpb}{x_{\mathrm{fp}(B)}}
\newcommand{\ycolli}{Y_{\mathrm{colli}}}
\newcommand{\lumexp}{L_{\mathrm{exp}}}
\newcommand{\numexp}{N_{\mathrm{exp}}}
\newcommand{\lumsim}{L_{\mathrm{sim}}}
\newcommand{\numsim}{N_{\mathrm{sim}}}
\newcommand{\domegasim}{\Delta \Omega_{\mathrm{sim}}}
\newcommand{\fnorm}{F_{\mathrm{norm}}}
\newcommand{\koptimal}{K_{\mathrm{optimal}}}
\begin{document}
\preprint{APS/123-QED}
%
%
%
%
\title{
Measurement of the Generalized Polarizabilities of the Proton at Intermediate $Q^2$ }
%
%
%
%
%
%
\def\kph{\affiliation{Institut f\"ur Kernphysik, Johannes  Gutenberg-Universit\"at Mainz, D-55099 Mainz, Germany}}
\def\clermont{\affiliation{Universit\'e Clermont Auvergne, CNRS/IN2P3, LPC, F-63000 Clermont-Ferrand, France}}
\def\zagreb{\affiliation{Department of Physics, Faculty of Science, University of Zagreb, 10000   Zagreb, Croatia}}
\def\stefan{\affiliation{Jo\v zef Stefan Institute, SI-1000 Ljubljana, Slovenia}}
\def\unil{\affiliation{Faculty of Mathematics and Physics, University of Ljubljana, SI-1000 Ljubljana,  Slovenia}}  
\def\uniwash{\affiliation{Institute  for  Nuclear  Studies,  Department  of  Physics, The  George  Washington  University,  Washington  DC  20052,  USA}}
\def\utemple{\affiliation{Temple University, Philadelphia, PA 19122, USA}}
\def\mitlns{\affiliation{Laboratory for Nuclear Science, Massachussetts Institute of Technology, Cambridge, MA 02139, USA}}
\def\stonybrook{\affiliation{Department of Physics and Astronomy, Stony Brook University, SUNY, Stony Brook, NY 11794-3800, USA}}
\def\rikenbnl{\affiliation{RIKEN BNL Research Center, Upton, NY 11973-5000, USA}}
\def\unewmexico{\affiliation{New Mexico State University, Las Cruces, NM 88003, USA}}
\def\umississippi{\affiliation{Mississippi State University, Starkville, MS 39762, USA}}
%
\author{H.~Fonvieille}\email[]{helene.fonvieille@clermont.in2p3.fr}\clermont
\author{J.~Beri\v{c}i\v{c}}\stefan
\author{L.~Correa}\clermont\kph  
\author{M.~Benali}\clermont
\author{P.~Achenbach}\kph      
\author{C.~Ayerbe~Gayoso}\umississippi
\author{J.C.~Bernauer}\stonybrook\rikenbnl 
\author{A.~Blomberg}\utemple 
\author{R.~B\"ohm}\kph
\author{D.~Bosnar}\zagreb      
\author{L.~Debenjak}\stefan  
\author{A.~Denig}\kph
\author{M.O.~Distler}\kph
\author{E.J.~Downie}\uniwash        
\author{A.~Esser}\kph           
\author{I.~Fri\v{s}\v{c}i\'{c}}\mitlns
\author{S.~Kegel}\kph
\author{Y.~Kohl}\kph
\author{M.~Makek}\zagreb       
\author{H.~Merkel}\kph
\author{D.G.~Middleton}\kph
\author{M.~Mihovilovi\v{c}}\kph\stefan\unil
\author{U.~M\"uller}\kph
\author{L.~Nungesser}\kph
\author{M.~Paolone}\unewmexico 
\author{J.~Pochodzalla}\kph         
\author{S.~S\'anchez Majos}\kph
\author{B.S.~Schlimme}\kph
\author{M.~Schoth}\kph
\author{F.~Schulz}\kph
\author{C.~Sfienti}\kph 
\author{S.~\v{S}irca}\unil\stefan
\author{N.~Sparveris}\utemple 
\author{S.~\v{S}tajner}\stefan
\author{M.~Thiel}\kph
\author{A.~Tyukin}\kph
\author{A.~Weber}\kph       
\author{M.~Weinriefer}\kph
\collaboration{A1 Collaboration}
\date{\today}
\begin{abstract}
\begin{description}
\item[Background] Generalized polarizabilities (GPs) are important observables to describe the nucleon structure, and measurements of these observables are still scarce. \item[Purpose] This paper presents details of a virtual Compton scattering (VCS)  experiment, performed at the A1 setup at the Mainz Microtron by studying the  $\epg$ reaction.  The article focuses on selected aspects of the analysis. \item[Method] The experiment extracted the $\pllptte$ and $\plt$ structure functions, as well as the electric and magnetic GPs of the proton, at three new values of the four-momentum transfer squared $Q^2$: 0.10, 0.20 and 0.45 GeV$^2$. \item[Results] We emphasize the importance of the calibration of experimental parameters. 
The behavior of the measured $\epg$ cross section is presented and compared to the theory. A detailed investigation of the polarizability fits reveals part of their complexity, in connection with the higher-order terms of the low-energy expansion. \item[Conclusions] The presented aspects are elements which contribute to minimize the systematic uncertainties and improve the precision of the physics results. 
\end{description}
\end{abstract}
\maketitle

\section{\label{sec:introduction}Introduction}


Nucleon polarizabilities are fundamental observables which describe how the charge, magnetization and spin densities in the nucleon  are deformed when an external quasi-static electromagnetic field is applied. They can be accessed through the Compton scattering process $\rcsreactnucleon$, and owe their small magnitude~\cite{Tanabashi:2018oca} to the strong binding force of quantum chromodynamics. Polarizabilities extend to finite momentum transfer, by replacing the incoming real photon with a space-like virtual one ($\gamma^*$), of virtuality $Q^2$. This leads to the concept of generalized polarizabilities (GPs)~\cite{Arenhoevel:1974twc}, i.e., $Q^2$-dependent observables describing the spatial distribution of the polarization density in the composite system. Nucleon GPs are accessed in the virtual Compton scattering (VCS)  process $\vcsreactnucleon$, via the $\epgreactnucleon$ reaction. The associated theoretical framework was first established in Ref.~\cite{Guichon:1995pu}. Further developments~\cite{Drechsel:1997xv} led to six independent GPs at lowest order: two scalar ones,  the electric GP $\aeq$ and the magnetic GP $\bmq$, plus four spin GPs. These observables have a well-defined continuity to the polarizabilities in real Compton scattering (RCS) at $Q^2=0$.


The low-energy regime is defined by small values of the total energy $W$ in the $\gamma^* N$ center-of-mass (c.m.), typically below the pion production threshold, or slightly above it. In this regime, the photon electroproduction cross section is dominated by the so-called Bethe-Heitler(BH)+Born cross section, $\sigbhb$, that contains no polarizability effect and is entirely calculable in quantum electrodynamics. The effect of the GPs consists of a small deviation of the experimental $\epgreactnucleon$ cross section from $\sigbhb$.
The electric and magnetic GPs of the proton have been measured by several experiments, at various four-momentum transfers in the $Q^2$ range from 0.06 to 1.76 GeV$^2$ \cite{Roche:2000ng,dHose:2006bos,Laveissiere:2004nf,Fonvieille:2012cd,Bourgeois:2006js,Bourgeois:2011zz,Bensafa:2006wr,Sparveris:2008jx,Doria:2015dyx,Janssens:2008qe,Blomberg:2019caf,Bericic:2019faq}. GPs are extracted from $\epg$ cross sections by fitting methods based  either on the low-energy theorem~\cite{Guichon:1995pu} (low-energy expansion or ``LEX fit'') or the dispersion relation model for VCS~\cite{Pasquini:2001yy,Drechsel:2002ar} (``DR fit''). A more complete presentation can be found in the recent review~\cite{Fonvieille:2019eyf}.


Our VCS experiment has been conducted at the  Mainz Microtron (MAMI) at various times from 2011 to 2015, to perform new measurements of the electric and magnetic GPs of the proton in the intermediate $Q^2$ range. The results have been published in Ref.~\cite{Bericic:2019faq}, in terms of GPs and structure functions. 
The experiment was performed essentially below the pion production threshold, and GPs were extracted from the measurement of absolute $\epg$ cross sections, using the two fitting methods cited above. The aim of the present paper is to give complementary accounts of this experiment. After a brief review of the instrumental configuration (Sec.~\ref{sec:apparatus}), details of the analysis are provided, with a focus on calibration aspects (Sec.~\ref{sec:expanalysis}), photon electroproduction cross sections (Sec.~\ref{sec:csandnorm}) and polarizability fits (Sec.~\ref{sec:gpextraction}).

Cross-section data are available electronically as supplemental material to this article~\cite{crossvcsq2} and at arXiv.org in the source files.

\section{\label{sec:apparatus}The Experiment}

The experiment uses the unpolarized MAMI electron beam and the A1 setup with a 5 cm long liquid hydrogen target and the two high-resolution, small solid-angle magnetic spectrometers A and B in coincidence. We refer to Ref.~\cite{Blomqvist:1998xn} for a detailed description of the apparatus. The detector package comprises a set of vertical drift chambers and scintillators in each arm, plus a Cherenkov detector in the electron arm. The beam of intensity 5-15 $\mu$A is rastered on the target by 1-2 mm in both transverse directions. The instantaneous luminosity of the experiment reaches (0.6-1.8)$\times 10^{37}$ cm$^{-2}$/s.

The detected particles are the scattered electron and the outgoing proton of the  $\epg$ reaction. The event reconstruction yields the particles' four-momenta at the vertex, denoted by $\mathbf{k'}$ and $\mathbf{p'}$ for the final electron and final proton, respectively. The four-momentum of the missing particle (the outgoing photon), denoted by $\mathbf q'$, can then be reconstructed as  $\mathbf{q'=k+p-k'-p'}$, where $\mathbf{k}$ and $\mathbf{p}$ are the four-momenta of the incoming electron and the target proton, respectively. The missing mass squared, noted $\mxx =\mathbf{(q')^2}$, exhibits a clear peak corresponding to a single undetected photon, the so-called ``VCS events'' (cf. Fig.~\ref{fig:mx2-exp-sim}). The four-momentum of the virtual photon is $\mathbf{q=k-k'}$, with $Q^2 \equiv \mathbf{-q^2}$.


The experiment studies VCS at three yet unexplored values of $Q^2$: 0.10, 0.20 and 0.45 GeV$^2$.  The aim is twofold: to cover a rather large $Q^2$ range, while surrounding the point at  $Q^2= 0.33$ GeV$^2$ where previous measurements exist and are intriguing. 
An important variable for the design is the modulus of the three-momentum of the outgoing photon in the $(\gamma^* p)$ c.m., denoted by $\qpr$.  Two other main kinematical variables are the polar and azimuthal angles of the outgoing photon with respect to the virtual photon in the c.m., denoted by $\thcm$ and $\phicm$ respectively.

The low-energy theorem~\cite{Guichon:1995pu} is valid only below the pion production threshold, corresponding to $W= m_N + m_{\pi}$ and  $\qpr = 126$ MeV/$c$. Given the fact that the effect of the GPs in the cross section increases with $\qpr$, different energy regions are defined, according to their increasing sensitivity to the GPs:  ``low-$\qpr$'' ($\qpr < 50$ MeV/$c$) and ``high-$\qpr$'' ($\qpr > 50$ MeV/$c$). 
At each $Q^2$, three kinematical settings are chosen, each one with a different goal: 
$i$) a high-$\qpr$, out-of-plane setting (``OOP'') with large sensitivity to the electric GP,   
$ii$) a high-$\qpr$, in-plane setting (``INP'') with mixed sensitivity to the electric and magnetic GPs, and 
$iii$) a low-$\qpr$ setting  (``LOW'') with no sensitivity to the GPs but useful for normalization.
These settings are listed in Table~\ref{tab:kinematics}. Note that the $OOP_B$ angle of 8-9$^{\circ}$  in the  laboratory frame allows one to reach $\phicm=90^{\circ}$ in the c.m.  At each $Q^2$, the experiment is performed at a single value of the virtual photon polarization parameter $\epsilon$. The settings are designed to maximize this parameter, since large values of $\epsilon$ enhance the GP effect in the cross section.

\begin{table}
\caption{\label{tab:kinematics}
The main kinematical settings, in terms of beam energy $E_{\mathrm{beam}}$, spectrometer central momenta $P_A$ and $P_B$, spectrometer  angles relative to the beamline, $\theta_A$ and $\theta_B$, and the out-of-plane angle of spectrometer B ($OOP_B$). The scattered electron is detected in spectrometer B (resp. A) at $Q^2=$ 0.10 and 0.20 GeV$^2$ (resp. 0.45 GeV$^2$). A few complementary settings are also used as slight variants of these ones. 
}
\begin{ruledtabular}
  \begin{tabular}{ccccccc}
Setting & $E_{\mathrm{beam}}$ &  $P_A$  & $\theta_A$ &  $P_B$  & $\theta_B$ & $OOP_B$ \\
name  & (MeV) & (MeV/$c$) & ($^{\circ}$) &  (MeV/$c$) & ($^{\circ}$) & ($^{\circ}$) \\
\noalign{\smallskip}
\hline
\noalign{\smallskip}
\noalign{\smallskip}
\multicolumn{7}{c}{ $Q^2= 0.10$ GeV$^2$} \\
\noalign{\smallskip}
INP & 872 & 425 & 53.1 & 700  & 22.9 & 0 \\
OOP & 872 & 343 & 52.6 & 693  & 21.9 & 9.0 \\
LOW & 872 & 365 & 58.0 & 745  & 22.4 & 0 \\
\noalign{\smallskip}
\multicolumn{7}{c}{ $Q^2= 0.20$ GeV$^2$} \\
\noalign{\smallskip}
INP & 1002 & 580 & 51.5 & 766  & 30.4 & 0 \\
OOP & 1002 & 486 & 51.0 & 766  & 29.2 & 8.5 \\
LOW &  905 & 462 & 52.2 & 723  & 32.5 & 0 \\
\noalign{\smallskip}
\multicolumn{7}{c}{ $Q^2= 0.45$ GeV$^2$} \\
\noalign{\smallskip}
INP & 1034 & 650 & 51.2 & 634  & 32.7 & 0 \\
OOP & 1034 & 647 & 51.0 & 750  & 39.2 & 8.0 \\
LOW &  938 & 645 & 52.3 & 713  & 40.5 & 0 \\
  \end{tabular}
\end{ruledtabular}
\end{table}

High statistics are achieved in the experiment, with about 900k, 1100k and 300k VCS events recorded at $Q^2 =$ 0.10, 0.20 and 0.45 GeV$^2$, respectively. About one third of the statistics corresponds to low-$\qpr$ ($\qpr <$ 50 MeV/$c$) and is  used for absolute normalization (cf. Sec.~\ref{sec:csandnorm}). The remaining two thirds of events correspond to higher $\qpr$ and are used in the polarizability fits. 
The motivation for such high statistics is driven by considerations on the GP effect at backward $\thcm$ angles. This angular region is important because of its high sensitivity to the magnetic GP, via the structure function $\plt$. However, in this region the GP effect exhibits very rapid variations (cf. Fig.~\ref{fig:lex-dr-drminuslex}) and the LEX fit may not be applicable everywhere (cf. Sec.~\ref{sec:gpextraction}). To be able to include this region selectively in the fit, one needs a fine 2D-binning in $(\cthcm, \phicm)$, with reasonable statistics in each bin.

\section{\label{sec:expanalysis}Data analysis}

The experiment is quite demanding in terms of accuracy of the measured $\epg$ cross section. Indeed the effect of the GPs in the cross section is very small, ranging from a few percent to at most 15\%. The quality of the event reconstruction, the calibration of experimental parameters and the reliability of the simulation are key factors to minimize the systematic error and achieve competitive uncertainties of the physics results. 
A few-percent systematic error on the cross section induces non-negligible biases in the polarizability fits. We therefore aim  at a precision of  1\% on the knowledge of the solid angle, a goal that can be reached thanks to the excellent performances of the MAMI beam and  the A1 setup. Sections \ref{sec:event-recons} to \ref{sec:offsets-and-calib-params}  describe the steps towards this goal. Sections~\ref{sec:ana-cuts} and \ref{sec:event-rate} summarize the analysis cuts and the corrections to the event rate, while Sec.~\ref{sec:simulation} recalls a few features of the simulation.

\subsection{\label{sec:event-recons}Event reconstruction}

The event reconstruction is carried out by the A1 COLA software. In each spectrometer, vertical drift chambers provide a track of the detected particle in the focal plane, characterized by two transverse coordinates $( \xfp, \yfp )$ and two projected angles  $(\thfp, \phfp )$.  This track is transformed into variables of the particle at the target by using the spectrometer optics, described by the optical transfer matrix. One obtains four variables at the vertex: 
 the relative momentum $\delta = (P-P_{\mathrm{ref}})/P_{\mathrm{ref}}$ where $P_{\mathrm{ref}}$ is the reference momentum, the projected vertical and horizontal angles, $\theta_0$ and $ \phi_0$, respectively, as well as the transverse horizontal coordinate $y_0$ in the spectrometer frame. 
Some of this information is then coupled between the two spectrometers, to build more elaborate variables in the laboratory frame, such as the missing mass squared $\mxx$. The longitudinal coordinate of the  vertex, $\zve$, is obtained by intersecting the beam direction with the direction of the particle going into spectrometer B. This spectrometer is chosen for the vertex reconstruction, since its point-to-point focusing properties provide the optimal resolution in the $y_0$ coordinate.  $\zve$  depends therefore directly on $y_{0(B)}$.  
The transverse coordinates of the vertex, horizontal $(\yve )$ and vertical  $(\xve )$, are obtained solely from the beam position, and are formally equal to the instantaneous values of  the beam transverse positions $\ybeam$ and  $\xbeam$, respectively, corrected for the raster pattern. 
The time of coincidence between the two detected particles is formed by using the TDC information of the scintillators in each spectrometer.  Three other variables coupling the two spectrometers: $\qpr, \cthcm$ and $ \phicm$, are constructed for defining the 3D cross-section bins.

\subsection{\label{sec-exp-calib}Experimental calibration}

An important step of the analysis is the calibration of experimental parameters. After the raw calibration of detectors (documented, e.g., in Ref.~\cite{BericicPhD:2015}), a second level of calibration involves additional items, such as:  optical transfer matrix elements, various offsets in momenta, angles and positions, and a specific parameter describing the cryogenic deposit on the walls of the target cell (cf. Sec.~\ref{sec:offsets-and-calib-params}).
A major tool for judging the overall quality of the calibration is the missing mass squared $M_X^2$. It is sensitive to almost all parameters, but as a single variable it does not permit to adjust them all. Thus, different studies were developed off-line in order to fix all the experimental parameters. They are described in Sects.~\ref{sec:optical-studies} and \ref{sec:offsets-and-calib-params}.

\subsection{\label{sec:optical-studies}Optical studies}


A first study concerns the optical transfer matrices of the spectrometers. The work of Ref.~\cite{Blomqvist:1998xn} has established that, for spectrometer B,  a single set of optical coefficients can be used for dipole magnetic fields  up to 1.2 T, i.e., a reference momentum of 600 MeV/$c$. Ref.~\cite{Blomqvist:1998xn} also reports that, for spectrometer A, no field-dependent effects are seen up to 600 MeV/$c$, a value at which first indications of field saturation effects become visible. Above 600 MeV/$c$, the optical properties of the spectrometers may change increasingly due to magnetic saturation. 
In our experiment, spectrometer magnets are operated in the saturation region all the time for spectrometer B and about one third of the time for spectrometer A (cf. Table~\ref{tab:kinematics}). Calibration data taken during the experiment allow one to make some improvements with respect to the available spectrometer optics at high fields, namely for spectrometer B. This optimization work is outlined below.


Data taken with a stack of thin foils regularly spaced along the beam axis are used to optimize the optics in $y_{0(B)}$ at several central momenta between 635 and 765 MeV/$c$. The  $y_{0(B)}$ variable is of  special importance since it determines the longitudinal coordinate of the interaction point, $\zve$, on which  one of the main analysis cuts is applied (see Fig.~\ref{fig:zb-exp-sim} and Sec.~\ref{sec:ana-cuts}). Data with a sieve-slit collimator are taken to control the optics in the $(\theta_0,\phi_0)_{(B)}$ angles.

\begin{figure}[t]
\centerline{\includegraphics[width=\columnwidth]{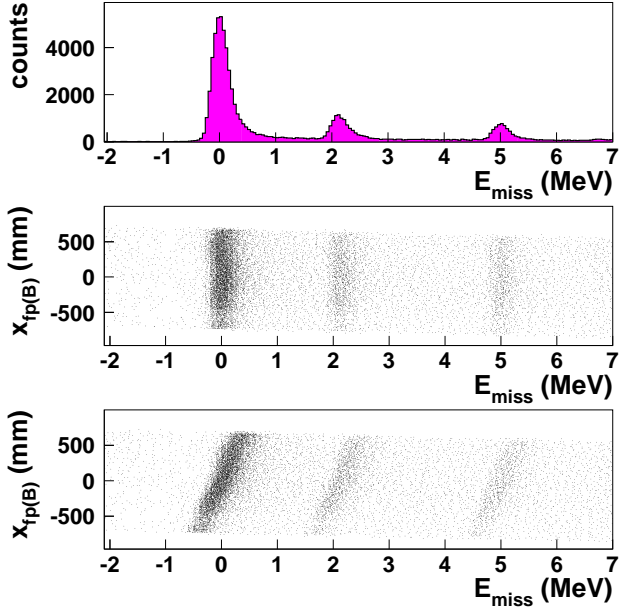}}
 \caption{\label{fig:nucl-peaks} (Color online) Top plot: nuclear levels in the reaction $^{12} C (e, e' p) X $ where the missing energy  $E_{miss}$ (corrected for kinematical broadening) represents the excitation energy of the $^{11}B$ nucleus. The peak FWHM is 0.30 MeV. Middle plot: the nuclear peaks versus the focal plane coordinate $\xfpb$, for a well-adjusted first-order coefficient  $( \delta \vert x )$ of spectrometer B. Bottom plot: the same thing for $( \delta \vert x )$ of spectrometer B decreased by 1\%. }
\end{figure}

For the relative momentum $\delta_{(B)}$, a few lowest-order optical coefficients can be partially adjusted on our $(e,e'p)$ coincidence data. The method is based on optimizing the width of the narrow peaks corresponding to nuclear levels in the missing energy spectrum. Such peaks originate from processes of the type $A(e,e'p)A$-1$^{(*)}$ and are observed in various calibration runs using a carbon target ($A=12$). They are also seen in ``VCS runs'' when the $\qpr$ variable, which actually corresponds to the missing energy, is small enough. In this last case, the nuclear $(e,e'p)$ events take place at the extreme ends of the cryotarget, where the beam crosses the walls of the cell and the cryogenic deposit. An example of nuclear peaks observed with a carbon target is given in  Fig.~\ref{fig:nucl-peaks}. The figure also illustrates the high sensitivity one can reach in the adjustment of the main first-order element $(\delta \vert x )$ (see  Eq.~(\ref{eq:optics}) for definition) with such events. Since this method uses both spectrometers at the same time, it relies on the good knowledge of the $\delta$-optics of one spectrometer, in order to tune the $\delta$-optics of the other spectrometer.

Based on the above adjustments, dedicated transfer matrices for spectrometer B have been devised and used at each central momentum setting. The optical transport is expressed by a polynomial expansion of the focal plane variables, given by the following set of equations (we adopt notations similar to Ref.~\cite{Blomqvist:1998xn}):
%
%
\begin{eqnarray}
\begin{array}{lll}
 \delta & = &  ( \delta \vert x ) \, \xfp +  
 ( \delta \vert \theta ) \, \thfp +  ...    \, ,     \\
 \theta_0 & = &  ( \theta \vert x ) \, \xfp +  
 ( \theta \vert \theta ) \, \thfp +  ...   \, ,  \\
 \phi_0 & = &  ( \phi \vert y ) \, \yfp +  
 ( \phi \vert \phi ) \, \phfp +   ...   \, ,  \\
 y_0 & = &  ( y \vert y ) \, \yfp +  
 ( y \vert \phi ) \, \phfp +   ...   \ .  \\
\label{eq:optics}
\end{array}
\end{eqnarray}

\begin{figure}[t]
\centerline{\includegraphics[width=\columnwidth]{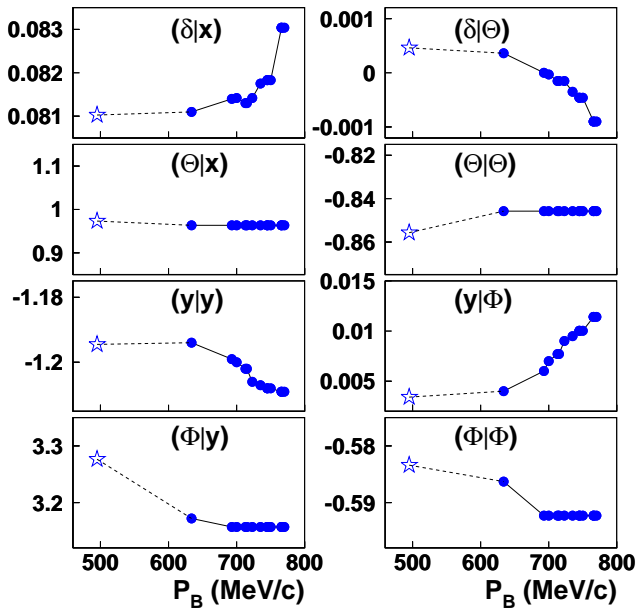}}
\caption{\label{fig:b-optics}
 (Color online) The eight first-order elements (cf. Eq.~(\ref{eq:optics})) of spectrometer B optics as a function of the central momentum, used in the VCS analysis. The experiment covers the region $P_B \in $ [634-770] MeV/$c$. The starred point indicates the non-saturated value at $P_B = 495$ MeV/$c$. The units in ordinate combine cm, mrad and percent.
}
\end{figure}

Here, only the eight first-order (and dominant) terms have been explicitly written out, and the dots indicate the series of higher-order terms, which are proportional to $ \xfp^i  \, \thfp^j \, \yfp^k  \, \phfp^l$. Figure~\ref{fig:b-optics} shows the eight first-order elements of the spectrometer B transfer matrices  used in the experiment, as a function of the central momentum $P_B$. 
Although  not deduced from a dedicated calibration campaign, and therefore not very accurate, they give an idea of the magnitude of the saturation effects in this spectrometer.  Overall, the observed variations are smooth versus $P_B$. The main terms: 
$ ( \delta \vert x ), ( \theta \vert \theta ), ( y \vert y )$ and $ ( \phi \vert \phi )$,  
 are only slightly affected by saturation effects, showing  at most a 2.5\% relative change in the displayed momentum range. For instance, the element $( y \vert y )$, which essentially gives the scale of the $y_{0(B)}$ reconstruction, is found to vary only by $\approx$ 1\% in the saturation region. However, ignoring this change would induce an error of up to 1\% on the scale of the target length, and hence a systematic error of similar size on the measured cross section. Other first-order terms in Fig.~\ref{fig:b-optics}, such as   $( \delta \vert \theta )$ or $ ( y \vert \phi )$,  show larger relative  variations, but their contribution is comparatively small.

For spectrometer A, the same optimization work has not been done, since available optics in the saturation region (at $P_A=645$ MeV/$c$) give essentially satisfactory results, in terms of sieve-slit reconstruction, $\mxx$ width or  nuclear peaks width. We just note that hints of saturation are observed for a central momentum $P_A$ as low as 580 MeV/$c$.

\subsection{\label{sec:offsets-and-calib-params}Offsets and other calibration parameters}


 
\begin{table*}
\caption{\label{tab:offsets}
Various parameters having a direct impact on the reconstructed missing mass squared and/or the simulated acceptance. Items are listed in the first column. The second column indicates the existence of a real-time measuring device, or the origin of the offset. The third column specifies the potential need for an off-line adjustment. How often the latter should be done is indicated in the fourth column. The different adjustment methods are numbered in the last column.
}
\begin{ruledtabular}
  \begin{tabular}{lllll}
Type of offset or & Source of information or & need to   & Time basis & Method \\
calibration constant &  measuring device & adjust?  &    &  \\
\noalign{\smallskip}
\hline
\noalign{\smallskip}
\noalign{\smallskip}
Beam energy $\ebeam$ & measured by MAMI & no &  &   \\
Spectrometer angles relative to the beam & on-line readout & no & &   \\
Spectrometer A central momentum $P_A$ & measured by NMR probe & no &  &    \\
Transverse beam position, horizontal $\ybeam$ & punctual screenshots & yes & per run & I \\
Offset in horizontal angles $\phi_{0(A)}$ and $\phi_{0(B)}$   & related to spectrom. optics & yes & once for all &  I \\
Offset in horiz. coordinates $y_{0(A)}$ and $y_{0(B)}$  & related to spectrom. optics & yes & once for all&  I \\
Cryotarget longitudinal centering $\ztarget$ & pre-experiment surveys & yes & per cooldown & I \\
Transverse beam position, vertical $\xbeam$ & punctual screenshots & yes & per run &  II \\
Offset in vertical angles $\theta_{0(A)}$ and $\theta_{0(B)}$  & related to spectrom. optics & yes & once for all &  III \\
Cryogenic deposit on target walls  $\efrost$  & none & yes & per run &  IV \\
Spectrometer B central momentum $P_B$ & measured by Hall probe & yes & per field setting &  IV \\
\noalign{\smallskip}
  \end{tabular}
\end{ruledtabular}
\end{table*}

Many parameters are continuously monitored on-line in order to ensure stable data taking conditions. While the AQUA program performs data acquisition, the MEZZO software performs the slow control of basically every instrumental device in the A1 Hall: magnets, detectors, cryotarget, beam delivery,  etc., and most of these items are known with high precision in real time. Table~\ref{tab:offsets} gives a list of the parameters that have an impact on either the particle reconstruction, the missing mass squared $\mxx$, or the acceptance as calculated by the simulation.
Some of these items do not need adjustment since they are measured with high precision: $\approx$ $10^{-4}$ relative for the beam energy $\ebeam$, $\approx$ 0.1 mr for the spectrometer angles, and  $<$ $10^{-4}$ relative for the central momentum $P_A$. The other items of Table~\ref{tab:offsets} potentially need to be adjusted, essentially by off-line re-calibrations. The corresponding methods, listed in Table~\ref{tab:offsets}, are outlined below.



Method I focuses on a set of variables pertaining to the horizontal plane, and treats them altogether. For convenience, time-independent offsets are introduced for the $y_{0(A)}$ and $y_{0(B)}$ coordinates, and for the $\phi_{0(A)}$ and $\phi_{0(B)}$ angles of the reconstructed particles. The longitudinal position of the center of the cryotarget along the beamline, $\ztarget$, is known only to a limited precision. Indeed, the target may slightly move when going from warm to cold state, with a degree of reproducibility that is unknown. We  therefore consider one adjustable value of  $\ztarget$ for each new establishment of the cold state. The beam position on the target is not continuously monitored during the experiment, but only inspected visually at discrete times, by inserting a scintillating Al$_2$O$_3$ screen. The $\ybeam$ parameter (averaged over the raster) is thus re-determined for each run.


 A global fit of these different parameters is realized, based on several constraints on reconstructed variables: 
$i$) the target center, $\ztarget$, must be the same when seen by both spectrometers A and B, and must be constant over given periods of time; 
$ii$) the $Z$-position of the thin carbon target used in calibration runs  must be as close as possible to zero, to agree with precise pre-experiment surveys; 
$iii$) the edges of the entrance collimators must display a left-right symmetry in their positioning. Indeed, each collimator is centered by construction on the spectrometer's optical axis. The variable allowing this test is the reconstructed impact coordinate at the collimator plane: $\ycolli = y_0 + D  \tan  \phi_0 $,  where $D$ is the target-to-collimator distance.

As a result of this global optimization, performed on the entire data set, the center of the cryotarget is found to be shifted upstream along the beamline, by 1.4 mm  to 3.3 mm depending on the data taking period. This knowledge serves as an input to the simulation. The beam horizontal position is found to be very stable in time, with excursions smaller than  $\pm$ 1 mm relative to the nominal setpoint. Incidentally, this study also allows to quantify potential (horizontal) mispointings of spectrometer B  when the latter, weighing 2000 kN, is moved out-of-plane. In these uplifted configurations, and within the precision of the method, we observe no extra-offset in $\phi_{0(B)}$, and an extra-offset in $y_{0(B)}$ in the range (0.5-0.9) mm. These very small values testify to the remarkable stability of the spectrometer's mechanical alignment during out-of-plane motions.


Method II allows to adjust the vertical beam position  $\xbeam$ in-between the daily visual inspections. It uses the fact that variations in $\xbeam$  induce visible shifts in the sharp edges of the $\theta_{0(A)}$ distribution (the vertical angle of the particle), due to the very small target-to-collimator distance (0.56 m) in spectrometer A. Fitting the centroid of the $\theta_{0(A)}$ spectrum for each run provides an efficient follow-up of the $\xbeam$ variations with time. Observed excursions with respect to the nominal setpoint do not exceed  $\pm$ 1mm.


The remaining methods, III and IV,  make use of the missing mass squared in VCS. The aim is to optimize the $\mxx$ photon peak, i.e., to center it on its nominal position and minimize its width. This peak width is representative of the resolution achieved by the apparatus. As already mentioned, the $\mxx$ variable is kinematically sensitive to all particles' momenta and angles, and to the thickness of the cryogenic deposit on the target walls. A wrong value of these parameters causes distortions of the $\mxx$ peak, which in turn allow for diagnostics on some global offsets.


Method III focuses on possible global offsets attached to the vertical angles  $\theta_{0(A)}$ and $\theta_{0(B)}$ of the reconstructed particles. The $\mxx$ optimization does not constrain both parameters, but only a linear combination of them, of the type $(P_A \sin \theta_{0(A)}  + P_B \sin \theta_{0(B)})$. The main finding is that the adjustment hints at a small but noticeable vertical misalignment with respect to an ideal setup. An offset is needed that de-centers the distribution of either the $\theta_{0(A)}$ angle or the $\theta_{0(B)}$ angle. In the absence of further identification of its origin, this misbalance is entirely attributed to the $\theta_{0(A)}$ angle, de-centering its distribution by about 3.2 mr for the settings at $Q^2$ = 0.10 and 0.20 GeV$^2$, and 0.6 mr for the settings at $Q^2=$  0.45 GeV$^2$.   
In the simulation, this departure from an ideal setup is reproduced by shifting the entrance collimator of spectrometer A by about 1.8 mm downwards for the settings at $Q^2 = $ 0.10 and 0.20 GeV$^2$, and 0.3 mm downwards for the settings at $Q^2 = $ 0.45 GeV$^2$.


Method IV determines the last two unknown parameters. The first one is related to the cryogenic deposit around the target cell, due to residual nitrogen, oxygen and water vapor present in the scattering chamber. This deposit varies with time in an unpredictable way, and affects the acceptance through particle energy losses. This extra-material is modeled in the  analysis codes by a uniform layer over the cell, leading to one single adjustable item: the layer thickness, $\efrost$, in g.cm$^{-2}$. The second parameter is the value of the central momentum in spectrometer B, $P_B$. It is measured with a rather limited accuracy (a few per mil or more) by a Hall probe,  and needs to be more finely determined at each new field setting.

The key variables to optimize these two parameters are the position and the width of the $\mxx$ photon peak. Contrarily to the previous methods based solely on experimental data, here the simulation is also used. One can then exploit the two most-sensitive features: the high sensitivity of the peak position to  $P_B$  in the experiment, and the high sensitivity of the peak width to $\efrost$ in the simulation. We note in passing that, apart from the cryogenic deposit, all other sources contributing to the resolution in the simulation (cf. Sec.~\ref{sec:simulation}) are well constrained by other means.

This two-fold optimization leads to a unique solution in terms of  $P_B$ and the cryogenic deposit $\langle \efrost \rangle $ averaged over the  setting. As a last step, $\efrost$ is finely tuned run-per-run in the experimental sample, by requiring the position of the $\mxx$ photon peak to be stable in time. Overall, the thickness of the cryogenic deposit is found to vary in the range (0 - 0.1) g.cm$^{-2}$ throughout the whole data taking. The adjusted values of $P_B$ depart from the Hall probe readings by $\approx$  a few per mil, which is consistent with the expected accuracy of the measuring device.

\begin{figure*}[t]
\centerline{\includegraphics[width=17cm]{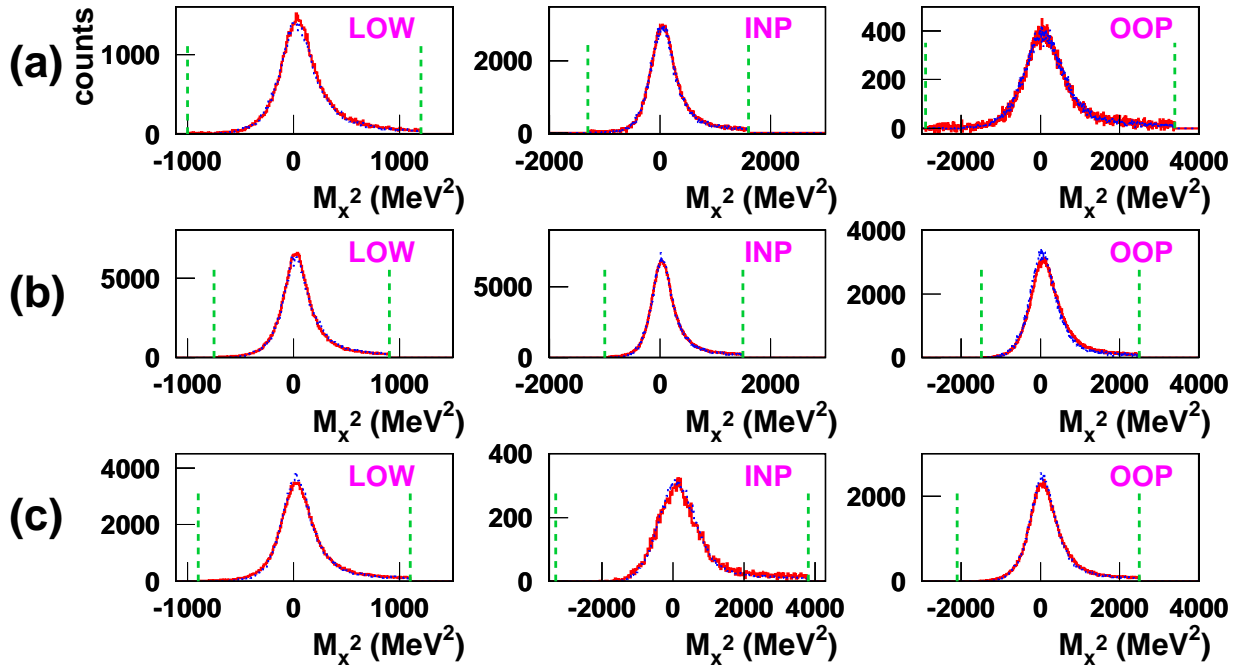}}
 \caption{\label{fig:mx2-exp-sim}
 (Color online) The experimental (solid red) and simulated (dotted blue) distributions of the missing mass squared, for each type of setting. Plots in columns refers to LOW, INP and OOP settings, while rows (a), (b), (c)  refer to $Q^2 =  0.10, 0.20$ and $0.45$ GeV$^2$, respectively. All the analysis cuts are applied. Both the experimental and simulated distributions are normalized to the same luminosity (i.e., there is no free adjustment). The lower and upper cuts in $\mxx$ (vertical dashed green lines) correspond to $-$6 and +7 $\sigma$, where  $\sigma$ is the r.m.s. of the photon peak. 
}
\end{figure*}

The resulting $\mxx$ distributions of Fig.~\ref{fig:mx2-exp-sim} show the good level of agreement obtained between the experiment and the simulation. Depending on the setting, the photon peak is centered on values ranging from 20 to 100 MeV$^2$ and the optimized width is in the range (300-1300) MeV$^2$ (FWHM). On average, the simulation and the experiment agree  to $\approx \pm 10$  MeV$^2$ on the peak centering, and to $\approx \pm 20$ MeV$^2$ on the peak width. This good agreement is also verified locally in the VCS phase space.

As a conclusion to this section, a good calibration of all the mentioned parameters is important to get the correct experimental event rate, as well as a faithful simulation. The accuracy reached by the above methods is estimated to be below $\pm$ 0.5 mm on the beam position ($\xbeam$ and $\ybeam$) and on $y_0$ offsets, $\pm$ 0.5 mr on the offsets in the ($\theta_0, \phi_0$) angles, $\pm$ 0.3 MeV/$c$ on $P_B$ and $\pm$ 0.01 g.cm$^{-2}$ on $\langle \efrost \rangle$. 
Dedicated simulation studies show that, for each parameter varying within its quoted precision, the corresponding uncertainty, or systematic error on the integrated solid angle is in  most cases well below 1\% relative. The most crucial case is the knowledge of $\langle \efrost \rangle$ for the settings at $Q^2=0.10$ GeV$^2$. In these kinematics, the outgoing protons have the lowest momenta (kinetic energies of 70-90 MeV) and the simulated acceptance is very sensitive to the proton's energy loss through the layer of cryogenic deposit. This parameter has to be known to better than $\pm$ 0.01 g.cm$^{-2}$ in order to control the solid angle to $\pm$ 1\%.

For further insight, we refer to Fig.~\ref{fig:errsyst-piechart} in Sec.~\ref{sec:systerrors}, which shows the contribution of uncertainties in the calibration parameters to the systematic error on the physics observables. Nine such parameters are included; note that the $y_{0(A)}$ and $y_{0(B)}$ offsets are replaced by a single offset in $\ztarget$ (known to better than $\pm$ 0.5 mm). As can be seen from Fig.~\ref{fig:errsyst-piechart}, the outcome is rather complex and cannot be anticipated easily, apart from the decreasing importance of the $\efrost$ parameter (sector 8) when $Q^2$ increases. The results differ from one $Q^2$ to another; dominant calibration uncertainties come from $\efrost$ at $Q^2=0.10$ GeV$^2$, $\ztarget$ at $Q^2=0.20$ GeV$^2$ and $\xbeam$ at $Q^2=0.45$ GeV$^2$.

\subsection{\label{sec:ana-cuts}Analysis cuts}


The VCS sample is obtained from the experimental data by selecting the true coincidences via a timing cut, and essentially applying two main analysis cuts, in $\zve$ and $\mxx$.

The coincidence time spectrum exhibits a narrow peak, over a wide plateau formed by random events. The FWHM of the peak is in the range (0.8-1.7) ns. The true coincidences are kept in a window of $\pm$ 5 ns around the peak center, and the random coincidences are subtracted by using the side bands of the spectrum. The level of random events under the peak is usually very low, typically a few percent of the true coincidences (after having applied the two main analysis cuts). However, it still reaches 20-40\% for a few settings.

\begin{figure*}[t]
\centerline{\includegraphics[width=17cm]{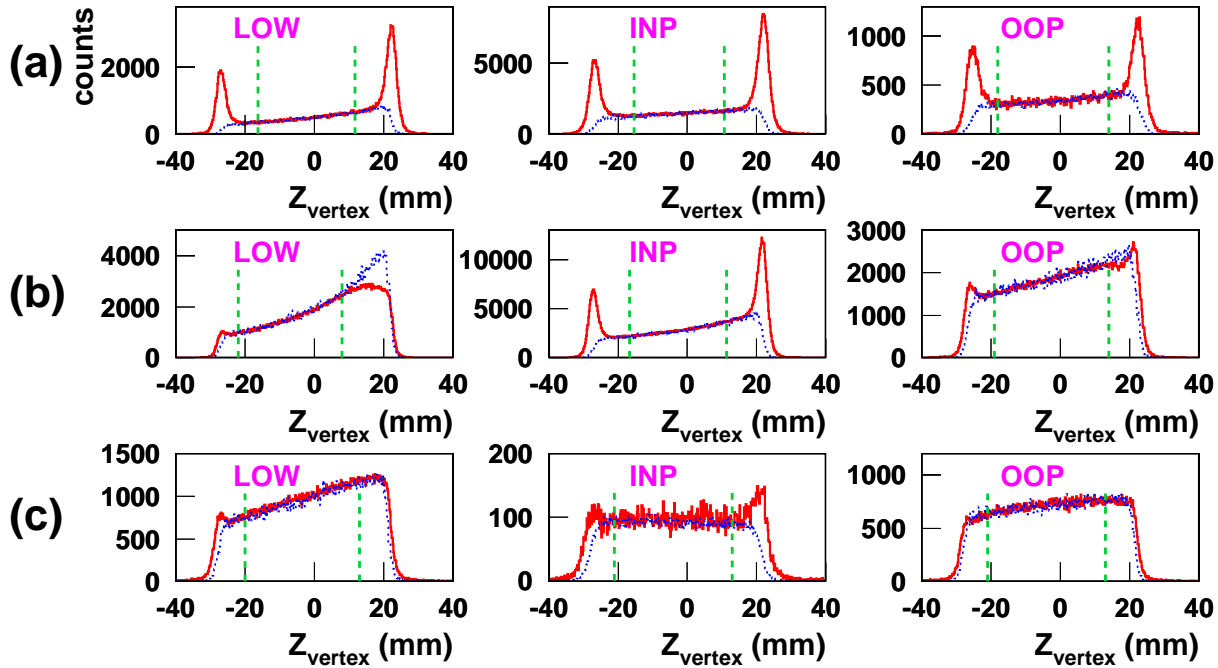}}
 \caption{\label{fig:zb-exp-sim}
 (Color online) The experimental (solid red) and simulated (dotted blue) distributions of the longitudinal vertex coordinate $\zve$, for each type of setting, with the same nomenclature for the plots as in the previous figure. The selected events are true coincidences within the $\mxx$ cut. Both the experimental and simulated distributions are normalized to the same luminosity. The useful part of the spectrum is the central region, delimited by the two vertical dashed green lines.
}
\end{figure*}

The need for a cut in $\zve$ is obvious from Fig.~\ref{fig:zb-exp-sim}, which compares the experiment and the simulation at the same level of cuts. While both event rates agree well in the central part of the target cell, they disagree at the extreme ends. For most settings, this region of the target shows an excess of experimental events relative to the simulation, due to  $(e,e'p)$ reactions on nuclei, not considered in the simulation. In one case (setting ``LOW'' at $Q^2=$ 0.20 GeV$^2$), a loss of experimental events, instead of an excess, is seen at the downstream end of the target. It may come from particles absorbed in the magnets for the events most close to elastic $ep \to ep$ kinematics. The cut (dashed vertical lines in Fig.~\ref{fig:zb-exp-sim}) selects the central part of the  $\zve$  spectrum,  reducing the usable target cell length to about 3 cm.

As the second main cut, events are required to be in the photon peak of the missing mass squared spectrum. The wide selection window around the peak center (cf. Fig.~\ref{fig:mx2-exp-sim}) allows one to include a large fraction of the radiative tail that develops on the positive-$\mxx$ side. These radiative events are well reproduced by the simulation.


The cut in  $\zve$ is the only one that eliminates a large fraction of VCS events. The cut in $\mxx$ just removes the distant part of the radiative tail. We now mention a few auxiliary cuts, which remove even smaller fractions of good events. 
Firstly, events are excluded when they are reconstructed far out of the nominal acceptance, either in the ($\theta_0, \phi_0$) angles, or in the impact point at the collimator, or in the relative momentum $\delta$. The selected window for $\delta$ is ($-$6,+16)\% in spectrometer A, and ($-$7,+7)\% in spectrometer B.
Secondly, for some settings a 2D-cut in the $(\mxx, \qpr)$ plane is designed to eliminate the few events at the most negative values of $\mxx$, which are seen in the experiment but not in the simulation. These events may come from $ep \to ep$ elastic scattering followed by particle rescattering inside the spectrometers.


After having applied all the cuts, one obtains a ``pure VCS'' experimental sample, very clean, as seen from Fig.~\ref{fig:mx2-exp-sim}.  In particular, there is no need for particle-identification (PID) cuts. This can be checked by testing the response of the PID detectors, i.e., the Cherenkov detector in the electron arm and the scintillators in the proton arm. At this stage of the analysis, there is extremely small trace, if any, of $\pi^-$ in the distribution of the Cherenkov signal, or $\pi^+$ in the distribution of scintillator  ADC signals.

\subsection{\label{sec:event-rate}Event rate corrections and luminosity}

The rate of experimental events, obtained after all cuts and the subtraction of random coincidences, is corrected for data acquisition deadtime. Since the scintillators are trigger elements, the event rate is also corrected for scintillator inefficiency. The latter is mapped in the $(x,y)$ coordinates in the scintillator planes, and found to be negligible almost everywhere, except in some localized regions at the overlap of the scintillator paddles. The efficiency of the vertical drift chambers is considered to be  100\% in all cases. At this stage, one obtains the number of experimental events $\numexp$ in each of the 3D cross-section bins. The precise measurement of the experimental luminosity $\lumexp$ relies on two inputs: the beam current, given by a fluxgate magnetometer, and the liquid hydrogen density, determined from pressure and temperature sensors. The continuous monitoring of these target parameters, together with the beam rastering,  ensure a very stable liquid hydrogen density.

\subsection{\label{sec:simulation}Simulation}

The acceptance, or solid angle $\Delta \Omega$ that is needed to determine the $\epg$ cross section, is too complex to be calculated by simple means. It requires the use of a simulation, as complete and faithful to the experiment as possible. We only summarize here the main features of the calculation of this acceptance, noted hereafter $\domegasim$. A more detailed description can be found in Ref.~\cite{Janssens:2006vx}.
The simulation only deals with $\epg$ events in the hydrogen volume of the cell, and does not consider any physical background or secondary processes.  $\domegasim$ is an ``effective'' and not purely geometrical solid angle, in the sense that all resolution effects are taken into account. The simulation includes the radiative effects which generate the tail in missing mass squared, and the effect of the cryogenic deposit around the target cell. Other sources of resolution consist in multiple Coulomb scattering, energy losses and straggling in the known materials, tracking errors in the focal plane and reconstruction errors at the target level. The description of the apparatus is based on the nominal characteristics (cf. Ref.~\cite{Blomqvist:1998xn}). Namely, the acceptance of the spectrometers is defined solely by the geometrical aperture of their entrance collimator, plus the nominal momentum acceptance. 
The simulation incorporates furthermore the results of the calibration described in Sec.~\ref{sec:expanalysis}, using setting-averaged parameter values. A simulated sample is obtained for each kinematical setting separately, together with its associated luminosity $\lumsim$. The simulated events are weighed by the realistic BH+Born cross section. Analysis cuts are then applied to the simulated sample in a way similar to the experiment.

\section{\label{sec:csandnorm}Cross sections and normalization}

\begin{table}
\caption{\label{tab:fnorm}
Results of the normalization test at each $Q^2$, using the data of the ``LOW'' settings (and their variants) at $\qpr=37.5$ MeV/$c$. The fitted value of the normalization factor is given in the third column, together with its statistical uncertainty obtained at ($\chi^2_{\mathrm{min}}+1$) (non-reduced $\chi^2$). The reduced $\chi^2$ of the fit and the number of degrees of freedom are given in the fourth and fifth columns, respectively.  The test uses the proton form factors parametrization of Ref.~\cite{Friedrich:2003iz} for calculating $\sigbhb$. 
}
\begin{ruledtabular}
  \begin{tabular}{lllll}
$Q^2$ & Setting & fitted $\fnorm$ &   $\chi^2$ & n.d.f. \\
\noalign{\smallskip}
\hline
\noalign{\smallskip}
\noalign{\smallskip}
 0.10 GeV$^2$ & LOW (I)     &   0.9856 $\pm$ 0.0063   &   1.22   &   400 \\
 0.10 GeV$^2$ & LOW (II)    &   1.0092 $\pm$ 0.0042   &   1.10   &   483 \\
 0.10 GeV$^2$ & LOW (III)   &   0.9704 $\pm$ 0.0029   &   1.10   &   600 \\
\noalign{\smallskip}
\noalign{\smallskip}
 0.20 GeV$^2$ & LOW (I)     &   0.9894 $\pm$ 0.0032   &   1.25   &   903 \\
 0.20 GeV$^2$ & LOW (II)    &   0.9885 $\pm$ 0.0034   &   1.10   &   817 \\
\noalign{\smallskip}
\noalign{\smallskip}
 0.45 GeV$^2$ & LOW         &   1.0173 $\pm$ 0.0041   &   1.11   &   712 \\
\noalign{\smallskip}
  \end{tabular}
\end{ruledtabular}
\end{table}

The  $\epg$ absolute cross section is the five-fold quantity  $d^5 \sigma_\mathrm{exp} / ( d E_e' d \Omega_e' d \cthcm d \phicm ) $, denoted hereafter by $\sigexp$.  $dE_e'$ and $ d \Omega_e'$ are the differential energy and solid angle of the scattered electron in the laboratory frame, while ($d \cthcm d \phicm$) is the differential solid angle of the emitted photon in the c.m. At each of the three $Q^2$, $\sigexp$ is determined at fixed $\qcm$ and fixed $\epsilon$, in a three-dimensional binning in the variables $( \qpr, \cthcm, \phicm )$. One obtains  $\sigexp (i)$ in each bin $i$ as  (cf. Ref.~\cite{Janssens:2006vx}):
%
%
\begin{eqnarray}
\sigexp (i)  \ = \  
\frac { \numexp(i)}{ \lumexp } 
\cdot
{\bigg [ } \,
\frac { \lumsim }{ \numsim(i) } 
\cdot
\sigbhb (i)
\, {\bigg ]}  \ , 
\end{eqnarray}
%
%
where $\numexp(i)$ is the number of experimental events in the bin, and $\numsim (i)$ the weighed sum of simulated events in this bin. The cross section $\sigbhb (i)$ is evaluated at the center of each bin, and the bracket represents the inverse of the five-fold solid angle $\domegasim$.

The chosen bin size is small: 25 MeV/$c$ in $\qpr$, 0.05 in $\cthcm$ and 10$^{\circ}$ in $\phicm$, allowing one to follow the rapidly varying effect of the GPs in this 3D phase space. As a result, many cross-section points are generated, of the order of a thousand at each $Q^2$. Our measured cross-section data are provided as supplemental material to this article~\cite{crossvcsq2}.

As explained in Ref.~\cite{Bericic:2019faq},  the final normalization of the experiment is based on the very low-$\qpr$ data, here $\qpr=37.5$ MeV/$c$. The method uses the fact that, at these low final photon energies, the measured cross section must coincide with the theoretical one, composed of the BH+Born cross section plus a very small GP effect ($<$1\%). $\sigbhb$ is entirely calculable when one makes a choice for the electric and magnetic form factors of the proton, $G_E^p(Q^2)$ and $G_M^p(Q^2)$. Here and in all the following, the form-factor parametrization of Ref.~\cite{Friedrich:2003iz} is used. The comparison of the experimental and the theoretical cross sections at low $\qpr$  is then realized by a $\chi^2$-minimization, in which the fitted parameter is the global normalization factor $\fnorm$ to apply to $\sigexp$.
As shown in Table~\ref{tab:fnorm}, we obtain in all cases a very good fit (reduced $\chi^2$ of $\approx$ 1, for about 400 to 900 data points involved) and a normalization factor $\fnorm$ very close to 1.00, within $\approx$ 1-2\%. It is an important test that confirms the consistency of all the prior analysis steps.

If one uses another parametrization of the proton form factors, i.e., other values of $G_E^p(Q^2)$ and $G_M^p(Q^2)$, the normalization factors of Table~\ref{tab:fnorm} may change. However, the physics results of the experiment, i.e., the fitted GPs and structure functions,  remain essentially unchanged, as long as the same form factor choice is used for the normalization of $\sigexp$ and for the polarizability fits (see Ref.~\cite{Fonvieille:2019eyf} for more details).

\begin{figure}
\centerline{\includegraphics[width=\columnwidth]{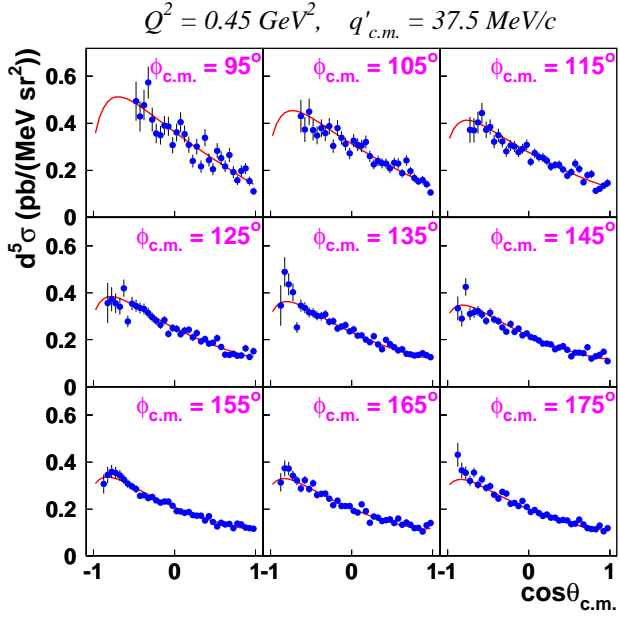}}
 \caption{\label{fig:cs-q2-05-lowqprim}
 (Color online) An example of the measured cross section at $Q^2=0.45$ GeV$^2$ and $\qpr = 37.5$ MeV/$c$. The solid (red) curve is the BH+Born calculation. Error bars are statistical only.
}
\end{figure}
\begin{figure}
\centerline{\includegraphics[width=\columnwidth]{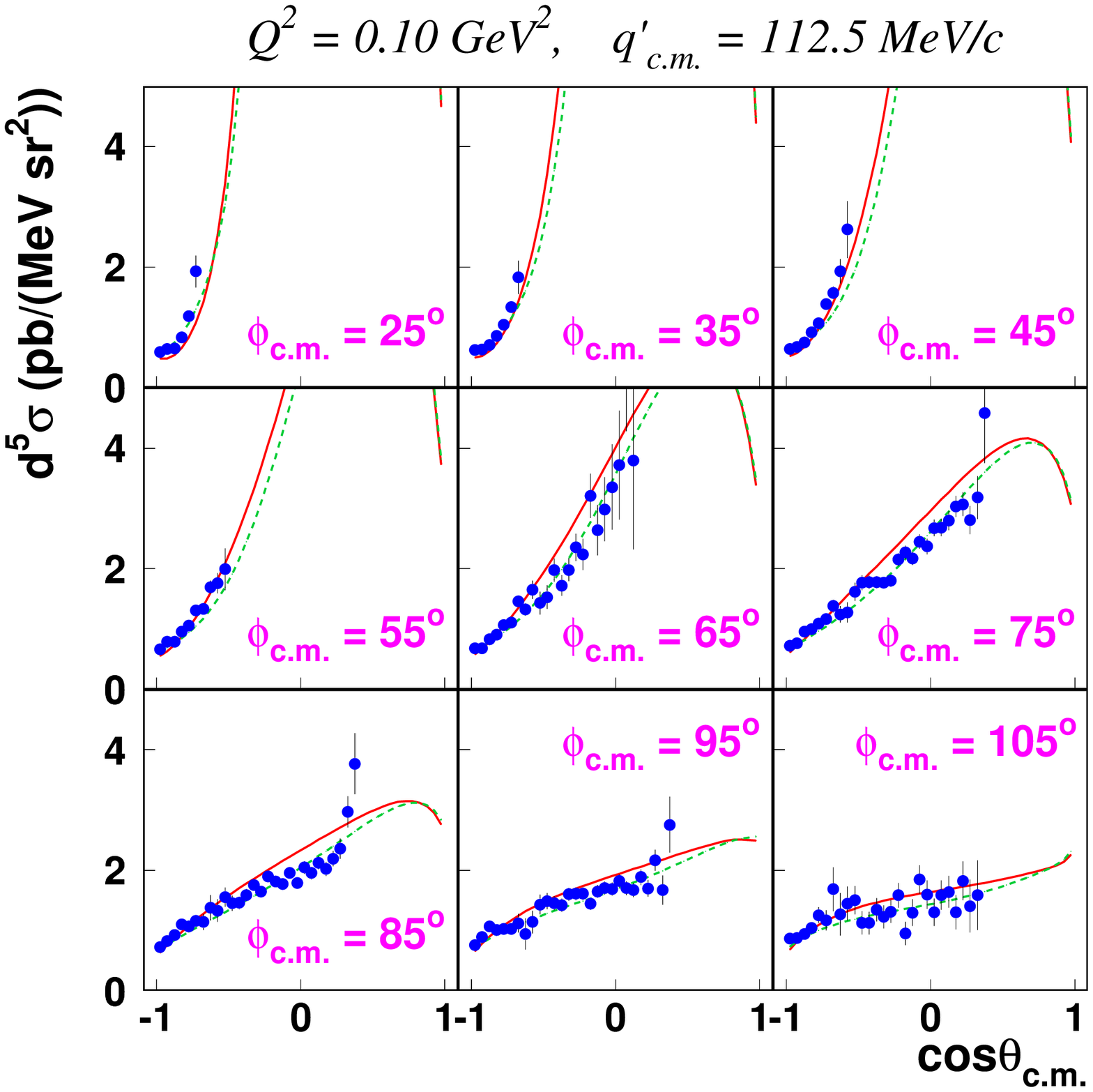}}
 \caption{\label{fig:cs-q2-01-highqprim}
 (Color online) An example of the measured cross section at $Q^2=0.10$ GeV$^2$ and $\qpr = 112.5$ MeV/$c$. The solid (red) curve is the BH+Born calculation and the dashed (green) curve includes in addition a first-order GP effect from the LEX. Error bars are statistical only.  
}
\end{figure}
\begin{figure}
\centerline{\includegraphics[width=\columnwidth]{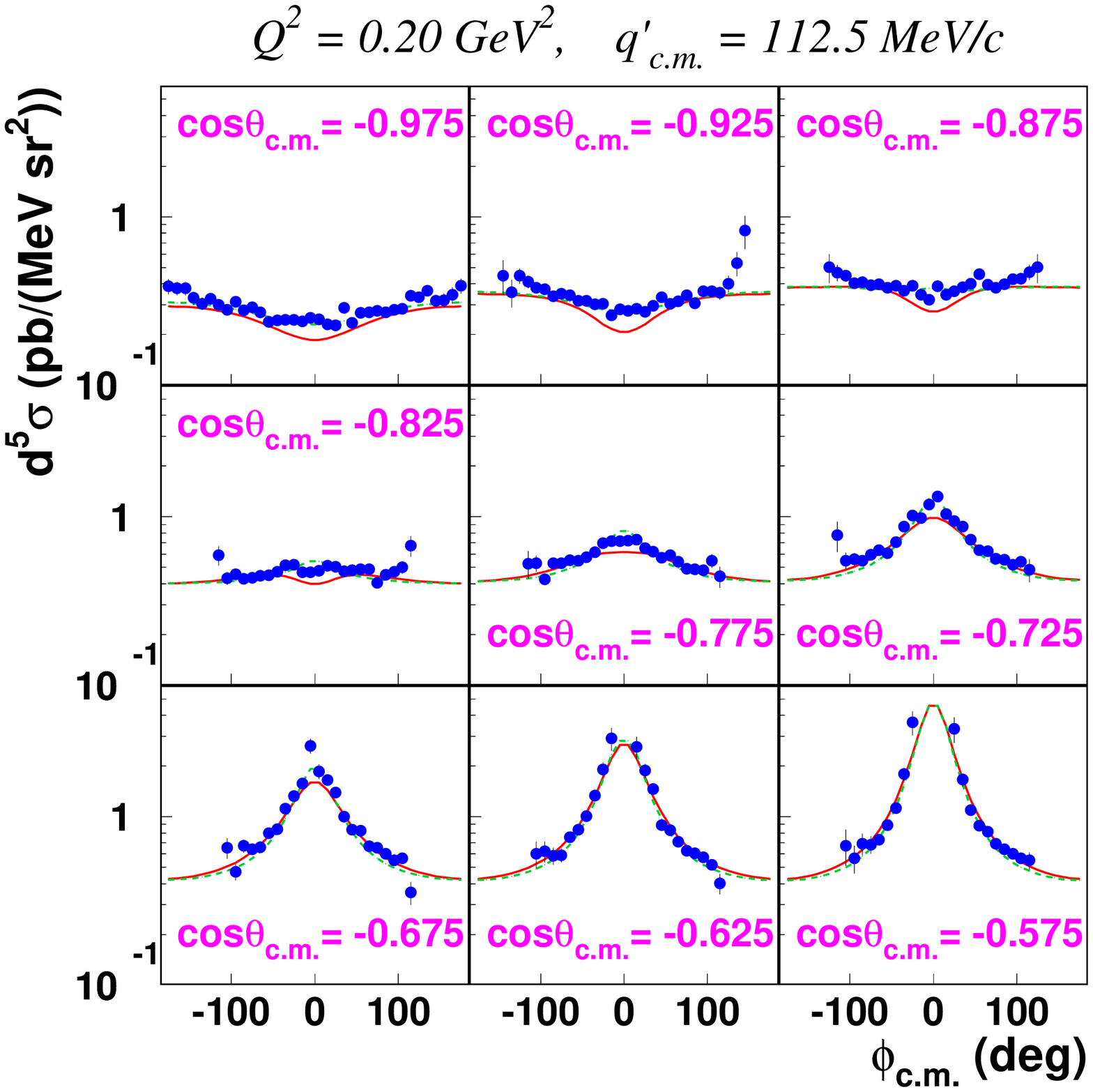}}
 \caption{\label{fig:cs-q2-02-highqprim}
 (Color online) An example of the measured cross section at $Q^2=0.20$ GeV$^2$ and $\qpr = 112.5$ MeV/$c$ (before symmetrization in $\phicm$). The solid (red) curve is the BH+Born calculation and the dashed (green) curve includes in addition a first-order GP effect from the LEX. Error bars are statistical only. 
}
\end{figure}
\begin{figure}
\centerline{\includegraphics[width=\columnwidth]{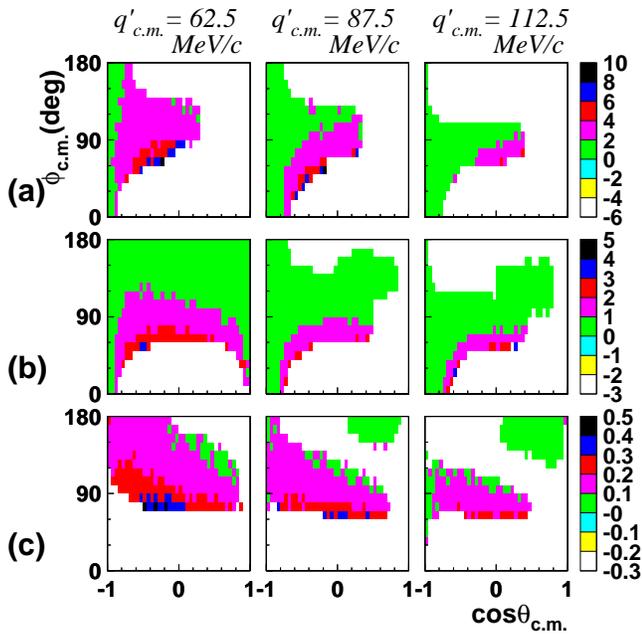}}
 \caption{\label{fig:cs-q2-all-2dplots}
 (Color online) A view of the VCS phase space covered by the experiment, in the $( \cthcm,\phicm )$ plane. Plots are made for three $\qpr$-bins, of central values 62.5, 87.5 and 112.5 MeV/$c$, from left to right. Rows (a), (b), (c) refer to $Q^2 =  0.10, 0.20$ and $0.45$ GeV$^2$, respectively. The content of each filled bin is the measured cross section, with the corresponding scale (one per $Q^2$) given at the right side of the figure, in pb/(MeV sr$^2$). 
}
\end{figure}

The next four figures show selected examples of our cross-section data.
Figure~\ref{fig:cs-q2-05-lowqprim} displays the low-$\qpr$ cross section  obtained at $Q^2=0.45$ GeV$^2$. As expected, no polarizability effect is observed here, and the measurement matches well the BH+Born cross section.  
Figures~\ref{fig:cs-q2-01-highqprim} and \ref{fig:cs-q2-02-highqprim} display the high-$\qpr$ data obtained at $Q^2=$ 0.10 and 0.20 GeV$^2$, respectively. On these figures one can discern  in some angular regions the small departure from $\sigbhb$ due to the GPs (the dashed green curves include the GP effect).
Figure~\ref{fig:cs-q2-02-highqprim} shows the quality of the symmetry of the cross section relative to $\phicm = 0 ^{\circ}$, a property that is required theoretically for an unpolarized experiment. Our final cross-section data~\cite{crossvcsq2} are subsequently symmetrized in $\phicm$. 
An overview of the experimental coverage in the   $(\cthcm, \phicm )$ phase space is given in Fig.~\ref{fig:cs-q2-all-2dplots}, for the three $\qpr$-bins  considered in the LEX fit. Each plot of this figure receives contributions from several kinematical settings, which are in some cases visible as isolated angular regions. Although most of the events are below the pion production threshold, the acceptance extends slightly beyond this limit. Namely, a small subset of cross-section values is obtained for the $\qpr$-bins [125-150] MeV/$c$ and [150-175] MeV/$c$ and will be considered in the DR fit.

\section{\label{sec:gpextraction}Extraction of the generalized polarizabilities}

We refer to Ref.~\cite{Fonvieille:2019eyf} for the detailed aspects of the formalism of VCS at low energy and methodologies for extracting the GPs from data. This section recalls the ingredients of the two fits using cross-section measurements below the pion production threshold: the LEX and DR fits. We further develop on an estimator of the higher-order terms of the low-energy expansion, which is used to make a detailed presentation of the fit results. Statistical and systematic errors are also discussed.

\subsection{\label{sec:theotools}Theoretical tools}

\begin{figure}[t]
\centerline{\includegraphics[width=\columnwidth]{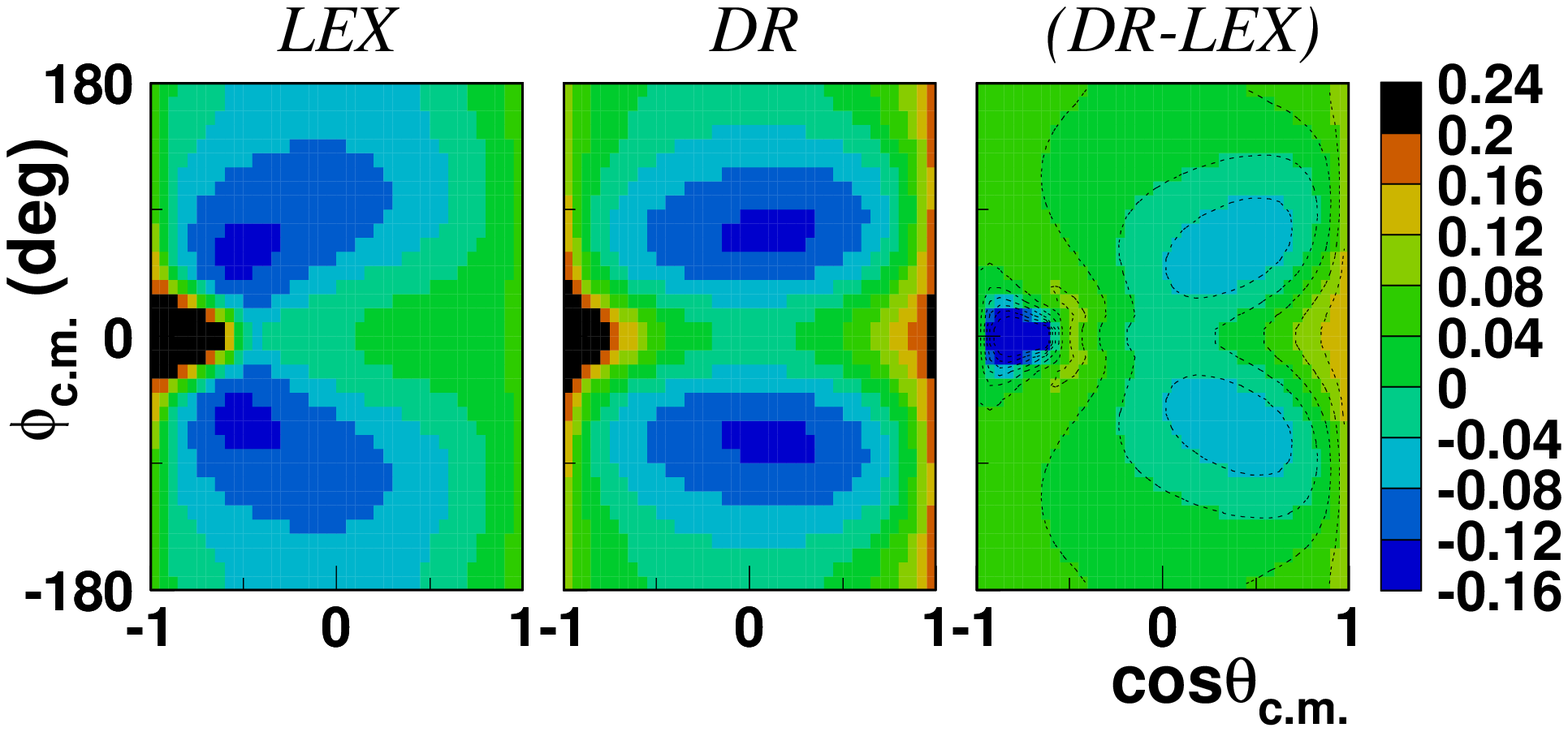}}
 \caption{\label{fig:lex-dr-drminuslex}
  (Color online) The GP effect in the 2D-plane ($\cthcm, \phicm$) at fixed $Q^2$ = 0.20 GeV$^2$, $\qpr$ = 110 MeV/$c$ and  $\epsilon$ = 0.9. Left: the GP effect from the LEX, defined as  $( \siglex - \sigbhb ) / \sigbhb $. Center: the GP effect from DR, defined as $ ( \sigdr - \sigbhb ) / \sigbhb$. Right: their difference, also equal to $( \sigdr - \siglex ) / \sigbhb $.  The calculation uses  $\pllptte = 15.5$ GeV$^{-2}$ and $\plt = -5.1$ GeV$^{-2}$ as input values.
}
\end{figure}


The LEX fit is based on the low-energy theorem~\cite{Guichon:1995pu}, a model-independent approach which expresses the $\epg$ cross section as: 
%
%
\begin{eqnarray}
\begin{array}{rll}
d \sigma & =  & \sigbhb +  ( \Phi \qpr ) \,  \Psi_0   +  \ohigher , \\
 \Psi_0  & =  & V_1  \, ( \pllptte ) + V_2 \, \plt  ,
\label{eq:lexformula}
\end{array}
\end{eqnarray}
%
%
where  $ \Phi \qpr, V_1, V_2$ are known kinematical factors. The three VCS response functions are the structure functions $\pll \propto \aeq$, $\plt \propto ( \bmq$ + spin GPs), and $\ptt  \propto$ spin GPs (see~\cite{Guichon:1998xv} for details). The  $\sigbhb$ cross section contains no polarizability effect and represents typically 90\% or more of the cross section below the pion production threshold. 
 $\Psi_0$ is the first-order polarizability term, and the quantity $[ \sigbhb +  ( \Phi \qpr ) \,  \Psi_0 ]$ will be denoted hereafter by $\siglex$. The higher-order terms $\ohigher$ are unknown and supposed to be small. They are neglected in the standard LEX fit, which therefore uses Eq.~(\ref{eq:lexformula}) in its truncated form without the $\ohigher$ term. 
A linear $\chi^2$-minimization compares $\sigexp$ with $\siglex$ and yields the two structure functions $\pllptte$ and $\plt$, at a given value of $Q^2$ and $\epsilon$. The electric and magnetic GPs are obtained only indirectly by this approach; an input from a model (here the DR model) is needed to subtract the spin-GP part of the fitted structure functions.  The LEX fit is performed for $\qpr$-bins below the pion threshold, in our case including the three bins [50-75], [75-100] and [100-125] MeV/$c$. The lowest $\qpr$-bin [25-50] MeV/$c$ serves essentially to fix the normalization and does not bring further constraint to the polarizability fit.


The DR fit is based on the dispersion relations model for VCS~\cite{Pasquini:2001yy,Drechsel:2002ar}, which has a wide range of applicability in energy, up to the $\Delta$ resonance region. In the DR formalism, the electric and magnetic GPs have an unconstrained part, which can be fitted to the experiment. $\aeq$ and $\bmq$ then become the two free parameters of the adjustment. $\sigexp$ is compared with the model cross section, $\sigdr$, calculated for all possible values of the free parameters, and $\aeq$ and $\bmq$  are fitted by a numerical $\chi^2$-minimization. The structure functions $\pllptte$ and $\plt$ are obtained from the scalar GPs in a straightforward way, by adding the contribution of the spin GPs, which is entirely fixed in the DR model. The DR fit uses the same $\qpr$-bins as the LEX fit, with the optional inclusion of bins at higher $\qpr$, above the pion production threshold.

\begin{figure}[t]
\centerline{\includegraphics[width=\columnwidth]{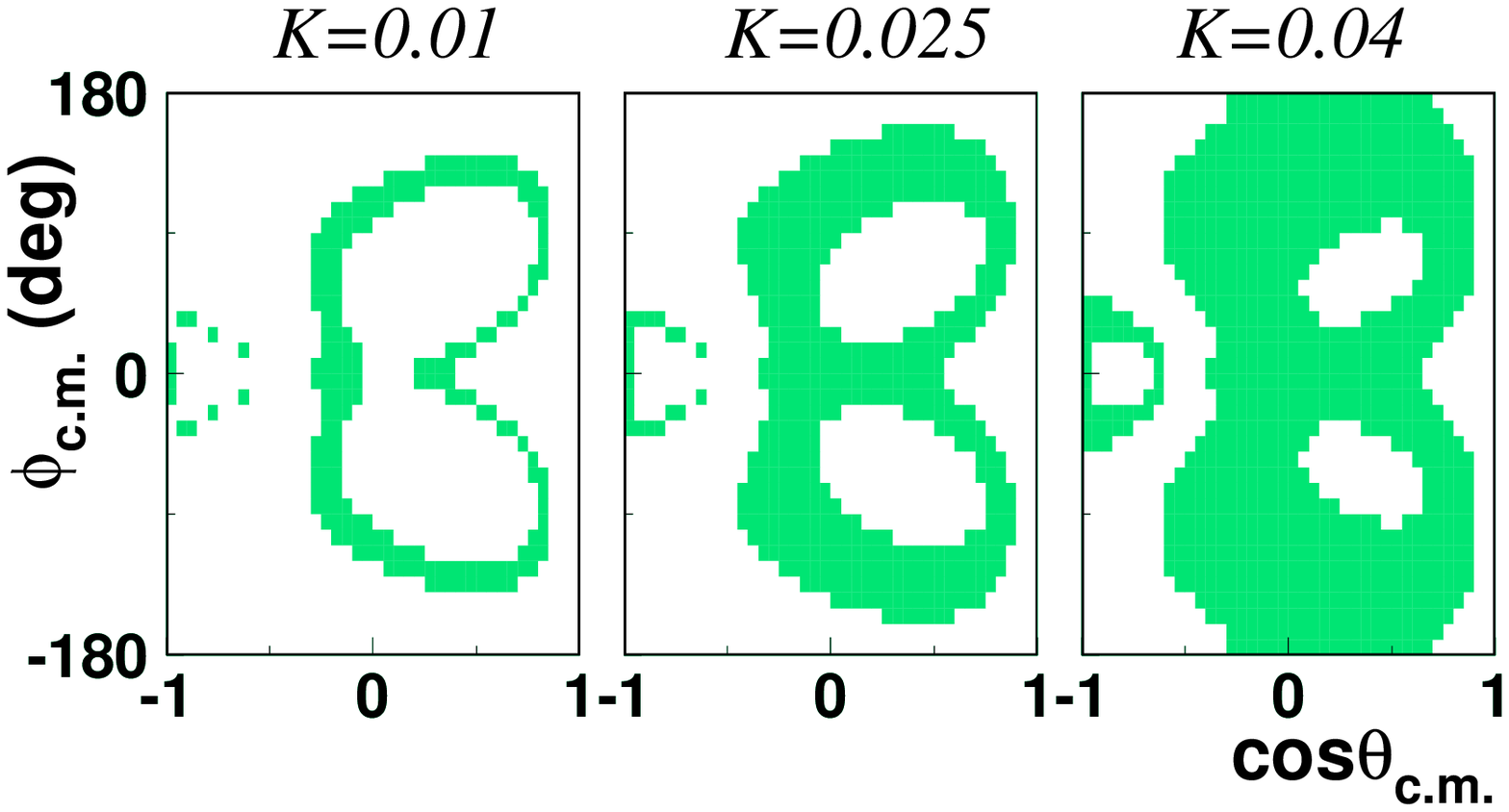}}
 \caption{\label{fig:three-maskthresholds}
 (Color online) Example of bin selection in the $(\cthcm, \phicm)$ plane, based on Eqs.~(\ref{eq:k-1}) and (\ref{eq:k-2}) (see text). Kinematics correspond to $Q^2 = 0.20 $ GeV$^2$, $\epsilon= 0.90$ and $\qpr = 112.5$ MeV/$c$. The plots show from left to right three increasing values of the cut threshold $K$, from 1\% to 4\%. Bins filled in green correspond to the condition $\ohigherdr \le K$. The calculation of $\ohigherdr$  uses  $\pllptte = 15.5$ GeV$^{-2}$ and $\plt = -5.1$ GeV$^{-2}$ as input values.
}
\end{figure}

\subsection{\label{sec:hoestimator}Higher-order estimator}

The LEX and DR fits are a priori very different, in the sense that $\siglex$ ignores the higher-order terms $\ohigher$, while $\sigdr$ includes by construction all orders in $\qpr$. When  these two fits are performed on the same data set, the appropriate comparison between their results is at the level of the structure functions $\pllptte$ and $\plt$, since these are the only direct outputs of the LEX fit. If both types of results agree, it is a strong indication that the higher-order terms $\ohigher$ of the LEX are indeed negligible. 
Among the various VCS experiments performed~\cite{Bourgeois:2011zz,Roche:2000ng,dHose:2006bos,Fonvieille:2012cd,Bericic:2019faq}, some find an agreement between the two types of fits, while others find a significant disagreement (see~\cite{Fonvieille:2019eyf} for more details). As a general statement, not much is known yet about these higher-order terms of the $\qpr$-expansion and their impact on the polarizability fits. In the present experiment, we have studied this question more systematically, using a novel method which is described in the remainder of this section.


Among its many advantages, the DR model can be utilized to provide an estimate of the higher-order terms of the LEX expansion. One just needs to calculate both theoretical cross sections, $\siglex$ and $\sigdr$, using the same input values of structure functions $\pllptte$ and $\plt$. Since  $\sigdr$ includes all orders in $\qpr$, the difference $( \sigdr - \siglex )$ is a measure of the higher-order terms $\ohigher$ of Eq.~(\ref{eq:lexformula}), as given by the DR model. Accordingly, we build the following dimensionless estimator:
%
%
\begin{eqnarray}
 \ohigherdr =  \frac{ \sigdr - \siglex }{  \sigbhb }  
\label{eq:k-1}
\end{eqnarray}
%
%
at each point in the VCS phase space. Figure~\ref{fig:lex-dr-drminuslex} shows an example of the GP effect calculated from the LEX, from the DR model, and their difference.

This (model-dependent) estimator has been used first in the design of the experiment~\cite{vcsq2proposal:2009}, to define kinematics where  $\ohigherdr$ is expected to be small. It is further employed in the analysis phase, to study the behavior of the LEX fit under varying conditions. More precisely, we perform the LEX fit of Eq.~(\ref{eq:lexformula})  in its truncated form, including a varying number of experimental bins, corresponding to gradually increased values of the $\ohigherdr$ estimator. This is realized by setting the condition 
%
%
\begin{eqnarray}
 \vert \ohigherdr \vert \le K \, , 
\label{eq:k-2}
\end{eqnarray}
%
%
and letting the threshold $K$ vary. An example of the accepted bins is given in Fig.~\ref{fig:three-maskthresholds}. In principle, this ``cursor'' for higher-order terms is not relevant for the DR fit, since the DR calculation is a priori valid in the whole VCS phase space. We have nevertheless performed the same study versus $K$ for the DR fit as well.


The $K$ parameter acts as a threshold for bin exclusion, or ``bin masking". A very tight cut, e.g., $K$ = 0.005, eliminates many bins in the $(\qpr, \cthcm, \phicm )$ phase space, mainly at high $\qpr$. In these conditions, The LEX and DR fits should give very similar results, since $\sigexp$ is compared to two model calculations, $\siglex$ and $\sigdr$, that almost do not differ.
As the cut threshold loosens, e.g., to $K$ = 0.02 or 0.03, more bins are included, larger differences between the two model calculations are allowed, and the LEX and DR fits may yield more different results.
At the largest value of the cut,  e.g., $K$ = 0.18 at  $Q^2$ = 0.20 GeV$^2$, all bins below the pion production threshold are included, and the LEX and DR fits become fully independent. This configuration is the one of the published LEX fits of all previous experiments~\cite{Roche:2000ng,Janssens:2008qe,Bourgeois:2011zz,Fonvieille:2012cd}.

\subsection{\label{sec:fitresults}Fit results}

\begin{figure}[t]
\centerline{\includegraphics[width=\columnwidth]{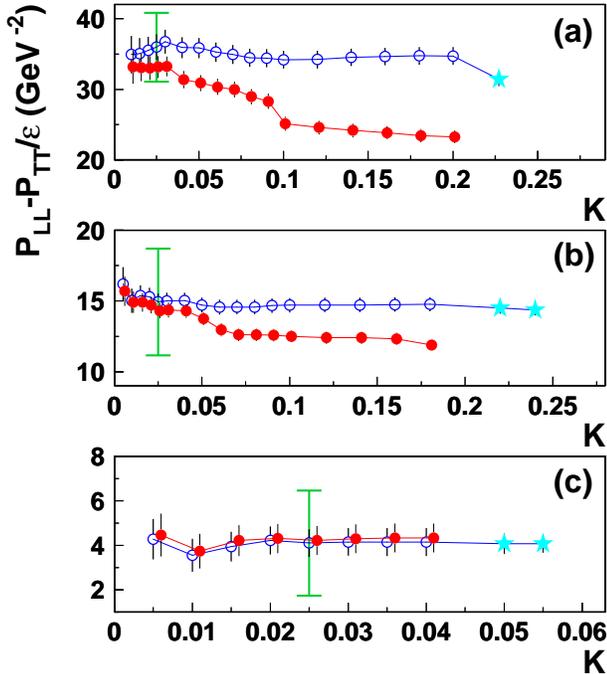}}
 \caption{\label{fig:sf1-versus-binmasking}
 (Color online) The behavior of the LEX fit (red filled circles) and the DR fit (blue open circles) as a function of the cut threshold  $K$ (see text) for the structure function $\pllptte$. Plots (a), (b), (c) refer to $Q^2 =  0.10, 0.20$ and $0.45$ GeV$^2$, respectively. In each plot, the rightmost filled (red) and open (blue) circles correspond to the inclusion of all data points in the $\qpr$-range (50-125) MeV/$c$ (i.e., the $K$-cut is inactive). The cyan starred points are placed at arbitrarily abscissa and refer to the DR fit with the inclusion of the $\qpr$-bin [125-150] MeV/$c$ (plot (a)) and additionally the $\qpr$-bin [150-175] MeV/$c$ (plots (b) and (c)). Error bars are statistical. The supplementary (green) error bar at $K=0.025$ represents the total systematic error, for our final choice of fit results.
}
\end{figure}
\begin{figure}[t]
\centerline{\includegraphics[width=\columnwidth]{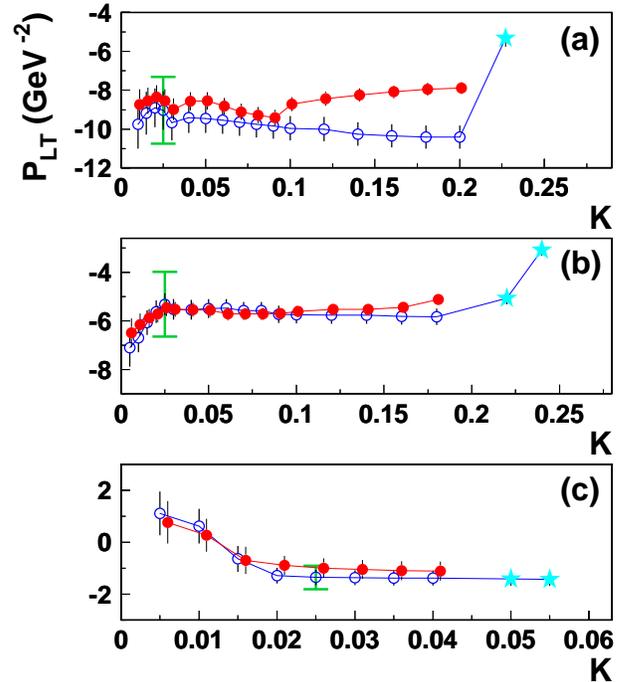}}
 \caption{\label{fig:sf2-versus-binmasking} 
 (Color online) The behavior of the fits for the structure function $\plt$.
Conventions are the same as in Fig.~\ref{fig:sf1-versus-binmasking}.
}
\end{figure}

Results of our fine scan in $K$ are shown in Figs.~\ref{fig:sf1-versus-binmasking} and \ref{fig:sf2-versus-binmasking} for the LEX and DR fits at each $Q^2$. At very small values of $K$, the two types of fits give very  similar results, as expected. When $K$ increases, the two fits tend to deviate, more or less quickly, indicating the effect of the higher-order terms $\ohigher$ that are neglected in the LEX fit. The divergence between the two types of fits versus $K$ is maximal for $Q^2=0.10$ GeV$^2$, and decreases when $Q^2$ increases. At $Q^2=0.45$ GeV$^2$, the two fits show no difference, suggesting  that the higher-order terms, as given by the DR model, are very small.

Another clear feature of Figs.~\ref{fig:sf1-versus-binmasking} and \ref{fig:sf2-versus-binmasking}  is the better stability of the DR fit versus $K$ relative to the LEX fit, in most cases. This demonstrates the good ability of DRs to evaluate the higher-order terms in $\qpr$ and to model the $\epg$ cross section over a large phase space. One notices a few localized exceptions to the stability of the DR fit versus $K$, for which possible origins can be invoked. At very small $K$ (Fig.~\ref{fig:sf2-versus-binmasking}, plots (b) and (c) for $K \le 0.02$), both the LEX and DR fits lack sensitivity to the GPs, due to the elimination of many high-$\qpr$ bins, and possible biases may arise. At the other end of the $K$ ``cursor'' (Fig.~\ref{fig:sf1-versus-binmasking}, starred point in plot (a), and Fig.~\ref{fig:sf2-versus-binmasking}, starred points in plots (a) and (b)), the added cross-section data above the pion production threshold correspond to acceptance edges,  where experimental systematics may be larger.


We now discuss how to choose the optimal value of $K$, for the  LEX fit with bin exclusion. For Eq.~(\ref{eq:lexformula}) to be valid in its truncated form, the higher-order terms should be small relative to the overall magnitude of the first-order GP effect, i.e., the $\Psi_0$ term. One is then led to choose small $K$ values, typically  $\koptimal <$ 3-4\%. A second qualitative argument is that higher-order terms will not bias the LEX fit significantly as long as their magnitude does not exceed the systematics of the experiment. This suggests $\koptimal \simeq 0.015$, corresponding to our  total systematic error of $\pm 1.5$\% on the measured cross sections (cf. Sec.~\ref{sec:systerrors}). Lastly, as mentioned above, the stability plateau for the DR fit in Figs.~\ref{fig:sf1-versus-binmasking} and \ref{fig:sf2-versus-binmasking} does not always start at the smallest value of $K$ but sometimes at $K \ge 0.02$.
Based on the above arguments, $\koptimal = 0.025$ is finally chosen, and considered as providing the most reliable LEX fit. This point is represented in Figs.~\ref{fig:sf1-versus-binmasking} and \ref{fig:sf2-versus-binmasking} with the attached total systematic error (thick solid green error bar).


In practice, the computation of  $\ohigherdr$ depends on input values for the structure functions, therefore the whole procedure (bin masking + polarizability fit) needs a few iterations. Figures~\ref{fig:sf1-versus-binmasking} and \ref{fig:sf2-versus-binmasking} are produced at the last iteration step. 
The results of both LEX and DR fits, obtained without bin masking and with bin masking at $K=0.025$, have been reported in Ref.~\cite{Bericic:2019faq}  for the structure functions and the scalar GPs. 
We also briefly report them here in Table~\ref{tab-results-lex-dr-fits} for the final choice of bin masking. Note the good quality of the fits, with reduced $\chi^2$  between 1.1 and 1.3, for $\approx$ 400 to 1000 degrees of freedom.

\begin{table*}
\caption{\label{tab-results-lex-dr-fits}
Results of the LEX and DR fits, obtained with bin masking at $K = 0.025$ (see text). The $\qpr$-bins cover the range (50,125) MeV/$c$. The first error is statistical. The second one is the total systematic error, whose sign indicates the correlation to the ($\pm$) sign of the overall normalization change. In the LEX part of the table, the GPs are obtained only indirectly, by subtracting from the structure functions the spin-GP contribution calculated by the DR model.
}
\begin{ruledtabular}
  \begin{tabular}{ccrrrrl}
\noalign{\smallskip}
$Q^2$   & $\epsilon$  &  $\pllptte$  &  $\plt$ & $\aeq$  & $\bmq$ & reduced $\chi^2$ \\
(GeV$^2$) & \ & ($\mathrm{GeV}^{-2}$) &  ($\mathrm{GeV}^{-2}$) &  (10$^{-4}$fm$^3$) &  (10$^{-4}$fm$^3$) & / n.d.f. \\
%
\noalign{\smallskip}
\hline
\noalign{\smallskip}
\noalign{\smallskip}
\multicolumn{7}{c}{LEX fit}   \\
\noalign{\smallskip}
\noalign{\smallskip}
%
%
0.10 & 0.91 &   33.15 $\pm$ 1.53 $\mp$ 4.53  &  $-$8.54  $\pm$ 0.60    $\mp$ 1.62  &  6.06  $\pm$ 0.30   $\mp$ 0.90 &  2.82   $\pm$ 0.23   $\pm$ 0.63 &  1.30/460 \\
0.20 & 0.85 &  14.57  $\pm$ 0.55   $\mp$ 3.47  & $-$5.37  $\pm$ 0.33   $\mp$ 1.25  & 3.02   $\pm$ 0.14  $\mp$ 0.87  &  2.01  $\pm$ 0.16   $\pm$ 0.61 & 1.29/1034  \\
0.45 & 0.63 & \, 4.21   $\pm$ 0.65   $\mp$ 2.24 & $-$1.00  $\pm$ 0.37   $\mp$ 0.50 & 0.92  $\pm$ 0.26  $\mp$ 0.92 &   0.19  $\pm$ 0.28   $\pm$ 0.38  & 1.17/820 \\ 
\noalign{\smallskip}
\noalign{\smallskip}
\multicolumn{7}{c}{DR fit}   \\
\noalign{\smallskip}
\noalign{\smallskip}
0.10 & 0.91  &  35.95 $\pm$ 1.80  $\mp$ 5.21 & \, $-$9.03  $\pm$  0.98  $\mp$ 1.82  & 6.60 $\pm$ 0.36 $\mp$ 1.03 &  3.02  $\pm$ 0.38  $\pm$ 0.72 &  1.34/460 \\
0.20 & 0.85 &  14.94 $\pm$ 0.60 $\mp$ 4.06 &  $-$5.31  $\pm$ 0.44 $\mp$ 1.40 &   3.11  $\pm$ 0.15  $\mp$ 1.02  &  1.98  $\pm$ 0.22  $\pm$ 0.68  & 1.31/1034  \\ 
0.45 & 0.63  &  \, 4.10  $\pm$ 0.62  $\mp$ 2.48 &  $-$1.36 $\pm$ 0.29  $\mp$ 0.40 & 0.87 $\pm$ 0.25  $\mp$ 1.01 & 0.47 $\pm$ 0.22  $\pm$ 0.30 & 1.14/820 \\
\noalign{\smallskip}
  \end{tabular}
\end{ruledtabular}
\end{table*}

We consider the results with bin masking (at $K=0.025$) as the final results of the experiment. However, one should keep in mind that they are a shorthand for a deeper complexity of the polarizability fits, of which some aspects have been explored and presented here.

\subsection{\label{sec:staterrors}Statistical errors}

Statistical errors on the physics observables are provided for each fit by the minimization itself, in which each term contributing to the $\chi^2$ is weighed by the statistical error on the measured cross section. The contour at ($\chi^2_{\mathrm{min}}+1$) (non-reduced $\chi^2$)  is used, corresponding to a confidence level of 70\% on  each  parameter separately. Error correlations between the two fitted parameters are small in all cases.

\subsection{\label{sec:systerrors}Systematic errors}

The dominant errors are the systematic ones. The normalization method based on the low-$\qpr$ data (cf. Sec.~\ref{sec:csandnorm}) helps to reduce them substantially, in the sense that all the global normalization uncertainties common to all settings, related for instance to the experimental luminosity or radiative corrections, are absorbed in the $\fnorm$  factor. 
However, residual normalization differences may still exist from setting to setting. They are  taken into account in a simplified way by considering an overall, intrinsic error of  $\pm$ 0.01  on  $\fnorm$. 
Another uncertainty comes from the calibration of experimental parameters and the solid angle calculation. Here again, the problem is simplified by considering the error globally, instead of possible point-to-point error correlations. The resulting uncertainty is estimated  to be $\pm$ 1\% on the cross section, relying on the work exposed in Sec.~\ref{sec:offsets-and-calib-params}. 
Lastly, another  $\pm$ 0.5\% uncertainty on the cross section is added as a way to take into account auxiliary, less significant sources of error, such as: possible non-uniformity of the virtual radiative correction factor in  the $(\qpr, \cthcm, \phicm )$ phase space, residual dependence of the physics results on the proton form factor choice, or versus the cut threshold $K$, etc.

\begin{figure}[!tbp]
\centerline{\includegraphics[width=\columnwidth]{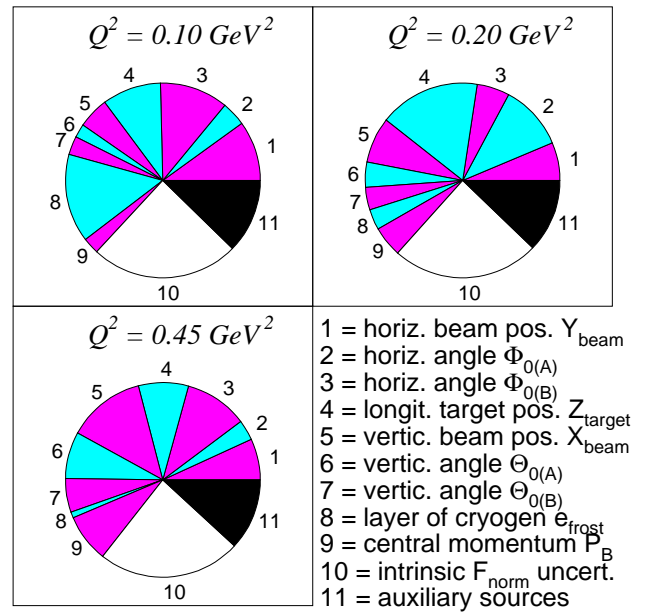}}
 \caption{\label{fig:errsyst-piechart} 
 (Color online) The detailed contributions to the total systematic error, for each $Q^2$. The pie chart represents the relative weights $w_i$ of each source of error ($i=1, ..., 11$). The indices $i=1, ..., 9$ correspond to uncertainties on the calibration parameters quoted in the concluding part of Sec.~\ref{sec:offsets-and-calib-params}. Here $i=10$ corresponds to the intrinsic uncertainty of $\fnorm$ and $i=11$ corresponds to the other auxiliary sources of error (see text, Sec.~\ref{sec:systerrors}). Due to our method of calculation, these charts apply equally well to the cross section, the structure functions and the GPs. 
Note that the total systematic error $\delta_{\mathrm{tot}}$ is given by the quadratic sum  $( \sum_{i=1}^{11} \delta_i^2 )^{1/2}  $, so that each partial error $\delta_i$ is given by  $ \delta_{\mathrm{tot}}  w_i ( \sum_{i=1}^{11} w_i^2 )^{-1/2} $  (with $\sum_{i=1}^{11} w_i = 1$).
}
\end{figure}

Figure~\ref{fig:errsyst-piechart} displays the systematic error budget at each $Q^2$, with the detailed contribution of each calibration parameter (corresponding to the nine colored sectors), as coming from simulation studies mentioned in  the concluding part of Sec.~\ref{sec:offsets-and-calib-params}. Summed quadratically, the eleven sources of error of Fig.~\ref{fig:errsyst-piechart} yield a total systematic error of $\pm$  1.5\% on the cross section. The latter is propagated to the physics results, by re-doing the polarizability fits with $\sigexp$ changed globally by $\pm$ 1.5\%.
This method yields errors on the physics results that are fully correlated in sign, due to the strong dependence of the two fitted parameters on the  $\fnorm$ factor. This ``one-shot'' method for obtaining the final systematic error is quick and efficient, but in some cases it is not realistic enough. We have tested the validity of this method by comparing it to more traditional means, such as performing various analyses with different calibrations, cut conditions, etc., and measuring the corresponding spread of the fitted results. 
On the one hand, the ``quick method'' works well at $Q^2=0.20$ GeV$^2$, as shown explicitly in Ref.~\cite{CorreaPhD:2016}, and is further assumed to work satisfactorily at $Q^2=0.10$ GeV$^2$, due to highly similar $(\qpr, \cthcm, \phicm )$ kinematics. On the other hand, this quick method  works only partly at $Q^2=0.45$ GeV$^2$, giving in particular an excessively small systematic error on $\plt$ (of $\pm$ 0.05 GeV$^{-2}$ for the LEX fit). The more traditional test of multiple analyses gives an error about ten times larger ($\pm$ 0.5 GeV$^{-2}$), which is clearly more realistic,  and chosen as the final value. Besides, both methods give a similar systematic error on $\pllptte$ at  $Q^2=0.45$ GeV$^2$.
 Although such disparities in the behavior of systematic errors are not fully traced, they could originate in differences of angular coverage in $(\cthcm, \phicm)$ versus $Q^2$, which induce differences in the weighing factors $V_1$ and $V_2$ of the low-energy theorem (cf. Eq.~(\ref{eq:lexformula})). We refer in particular to the angular coverage of the in-plane setting (``INP''), which corresponds to backward $\thcm$  angles at $Q^2=0.10$ and $0.20$ GeV$^2$, and to forward $\thcm$ angles at $Q^2=0.45$ GeV$^2$ (cf. in Fig.~\ref{fig:cs-q2-all-2dplots} the isolated region in green at $\cthcm > 0$ for $Q^2=0.45$ GeV$^2$).

\section{\label{sec:finalresults}Physics results and conclusions}

\begin{figure}[htbp]
\centerline{\includegraphics[width=\columnwidth]{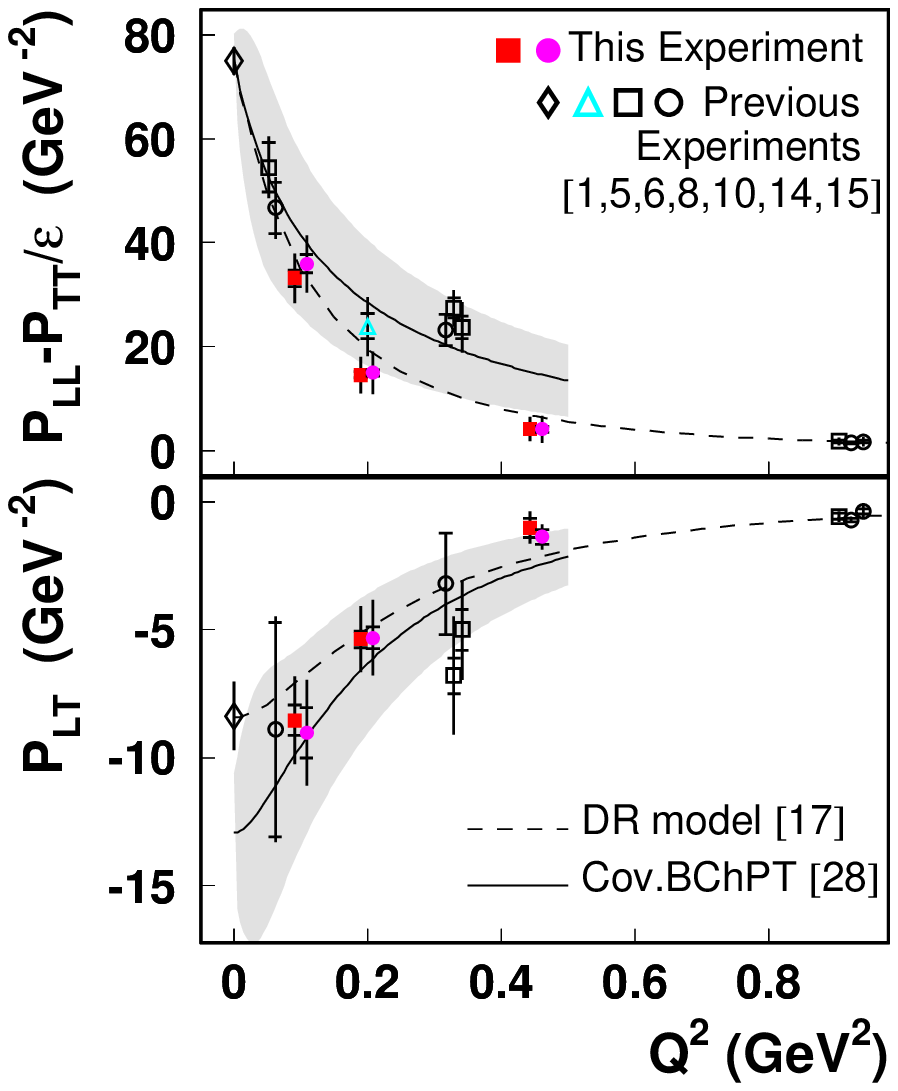}}
 \caption{\label{fig:final-sf-vs-q2}
 (Color online) The structure functions $\pllptte$ and $\plt$ of the proton (see text for details). Filled (magenta) circles and filled (red) squares at $Q^2$ = 0.10, 0.20 and 0.45 GeV$^2$  are from this experiment. Open circles and squares are from previous experiments at MIT-Bates~\cite{Bourgeois:2011zz} ($Q^2$ = 0.06 GeV$^2$), MAMI~\cite{Roche:2000ng,dHose:2006bos,Janssens:2008qe} ($Q^2$ = 0.33 GeV$^2$) and JLab~\cite{Fonvieille:2012cd} ($Q^2$ = 0.92 GeV$^2$). Open and filled circles correspond to DR analyses, while open and filled squares refer to LEX analyses. 
The triangular (cyan) point in the upper plot is from the recent measurement of the electric GP at $Q^2= 0.20$ GeV$^2$~\cite{Blomberg:2019caf}, converted to $\pllptte$ using the DR model. The RCS point ($\diamond$) is from Ref.~\cite{Tanabashi:2018oca}. The dashed curve is obtained using the DR model~\cite{Pasquini:2001yy}   with dipole mass parameters $\Lambda_{\alpha} = \Lambda_{\beta}$ = 0.7 GeV. The solid curve with its error band (shaded area) is from covariant BChPT~\cite{Lensky:2016nui}. Some data points are slightly shifted in abscissa for visibility. The inner and outer error bars are statistical and total, respectively.
}
\end{figure}

\begin{figure}[htbp]
\centerline{\includegraphics[width=\columnwidth]{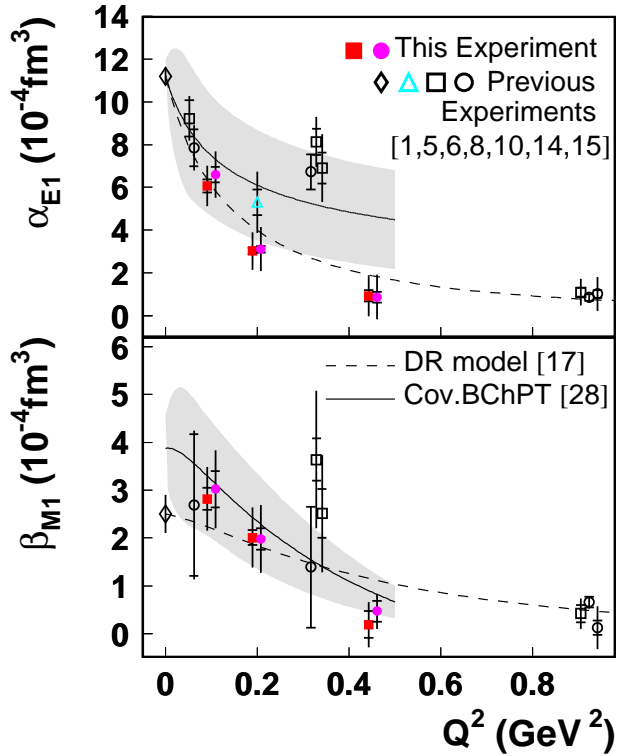}}
 \caption{\label{fig:final-ab-vs-q2}
 (Color online) The electric and magnetic GPs of the proton (top and bottom plots, respectively). The notations and conventions are  the same as in Fig.~\ref{fig:final-sf-vs-q2}.
}
\end{figure}

Our final results  are shown in Figs.~\ref{fig:final-sf-vs-q2} and \ref{fig:final-ab-vs-q2}, including the world data in terms of structure functions and scalar GPs of the proton. These results have been  discussed in Ref.~\cite{Bericic:2019faq} and in a broader context in Ref.~\cite{Fonvieille:2019eyf}, so we just summarize here the main findings. The present measurements provide important new insights into the $Q^2$-behavior of the VCS observables under study. A consistent and smooth behavior starts to emerge in the whole  $Q^2$ range from 0 to 1 GeV$^2$, with the exception of the existing data at $Q^2 = 0.33$ GeV$^2$ \cite{Roche:2000ng,dHose:2006bos,Janssens:2008qe}. The tension or lack of smoothness at this value of $Q^2$, observed especially for the $\pllptte$ structure function and the electric GP, remains presently unexplained and would require new investigations. 
A recently performed VCS experiment at Jefferson Lab~\cite{Sparveris:2016} is expected to shed light on this anomaly, by measuring the electric and magnetic GPs in the $Q^2$ range from 0.3 to 0.7 GeV$^2$.   
At  $Q^2 = 0.20$ GeV$^2$, results from the two most recent and independent experiments are shown for the electric GP and the $\pllptte$ structure function: the present measurement (filled circles and squares in the figures) and the one of Ref.~\cite{Blomberg:2019caf} (cyan triangular point). These two results show a rather good compatibility, although they involve different c.m. energy regimes: below the pion production threshold (our experiment) and the $\Delta$ resonance region ~\cite{Blomberg:2019caf}.

The DR model does not give a prediction of the electric and magnetic GPs. However, it uses a convenient parametrization of their $Q^2$-dependence, that allows to provide predictions for VCS observables. This is realized by assuming a single dipole behavior for the unconstrained part of the scalar GPs~\cite{Pasquini:2001yy,Drechsel:2002ar}. Namely, with dipole mass parameter values $\la = \lb = 0.7$ GeV (dashed curve in the figures), the DR model agrees well with the  $Q^2$-behavior suggested by the world data. 
The low-$Q^2$ data for the magnetic GP and the $\plt$ structure function show also good agreement with the recent covariant BChPT calculation of Ref.~\cite{Lensky:2016nui} (solid curve in the figures), despite the large theoretical uncertainty. Our experiment provides for the first time a precise measurement of $\bmq$ at very low $Q^2$ (0.10 GeV$^2$), strongly constraining the way the two large components, diamagnetic and paramagnetic, nearly cancel in this polarizability.


In conclusion, a new, high-statistics VCS experiment performed at MAMI has yielded precise measurements of the proton electric and magnetic GPs  at three yet unexplored values of $Q^2$. Although measurements of low-energy VCS observables are still rather scarce, they gradually improve in precision, as experiments are better designed and GP extraction methods become more mature. Examples along these lines have been given in this article. We have demonstrated how one can minimize systematic errors, by performing a careful experimental calibration and using the normalization constraint provided by low-$\qpr$ data. We have also shown how one can deepen the study of the polarizability fits themselves, in relation with the higher-order terms of the low-energy expansion. 
Nucleon GPs are valuable observables which bring specific constraints to models of nucleon structure. Improving their knowledge is a long-term challenge that will require inventive strategies for new measurements. The DR model,  with its unique advantages and evolutive capabilities, serves as a precious and reliable tool for designing and analyzing VCS experiments, and will help in pursuing further developments in the field.

\vskip 5mm
\begin{acknowledgments}
We wish to thank our theoretician colleagues Barbara Pasquini, Marc Vanderhaeghen, Vladimir Pascalutsa and Vadim Lensky for their support, and Vadim Lensky for providing the results of the covariant BChPT calculation. We gratefully acknowledge the MAMI-C accelerator group for the excellent beam quality. This work was supported by the Deutsche For\-schungsgemeinschaft with the Collaborative Research Center 1044, the Federal State of Rhineland-Palatinate and the French CNRS/IN2P3. Some of the authors would like to acknowledge the support by the Croatian Science Foundation under the project 8570. 
\end{acknowledgments}

\bibliographystyle{apsrev4-2.bst} 
\bibliography{vcsbiblio}
\end{document}